\documentclass[manuscript]{aastex}

\def\msun{{\rm\,M_\odot}}

\newcommand{\etal}{et al.\ }

\newcommand{\lya}{Ly$\alpha$ }

\def\etal   {{et~al.}\ }

\def\et{{\it et\thinspace al.}}    

\def\zsun{{\rm\,Z_\odot}}
\def\msun{{\rm\,M_\odot}}

\def\vol#1  {{{#1}{\rm,}\ }}

\def\etal{et al.\ }

\lefthead{Cen \etal} 
\righthead{The WHIM}

\begin{document}

\title{Where Are the Baryons? III: Non-Equilibrium Effects and Observables}

\author{
Renyue Cen$^{1}$ and Taotao Fang$^{2}$
} 
 
\footnotetext[1]{Princeton University Observatory, Princeton, NJ 08544;
 cen@astro.princeton.edu}
\footnotetext[2]{Department of Astronomy, University of California, Berkeley
, CA 94720; fangt@astro.berkeley.edu{\bf; {\sl Chandra} Fellow}}

\begin{abstract} 

Numerical simulations of the intergalactic medium have shown
that at the present epoch a significant fraction ($40-50\%$) of
the baryonic component should be found in the ($T\sim 10^6$K)
Warm-Hot Intergalactic Medium (WHIM) - with 
several recent observational lines of evidence indicating the validity 
of the prediction.
We here recompute the evolution of the WHIM
with the following major improvements:
(1) galactic superwind feedback processes from galaxy/star formation 
are explicitly included;
(2) major metal species (O V to O IX)
are computed explicitly in a non-Equilibrium way;
(3) mass and spatial dynamic ranges are larger by
a factor of 8 and 2, respectively, than in our previous simulations.
We find: (1) non-equilibrium calculations produce
significantly different results from ionization equilibrium calculations. 
(2) The abundance of O VI absorption lines based on
non-equilibrium simulations with galactic superwinds 
is in remarkably good agreement with latest observations,
implying the validity of our model,
while the predicted abundances for O VII and O VIII absorption lines
appear to be lower than observed but the observational errorbars
are currently very large.
The expected abundances for O VI (as well as \lya),
O VII and O VIII absorption systems 
are in the range $50-100$ per unit redshift at $EW=1$km/s decreasing to 
$10-20$ per unit redshift at $EW=10$km/s.
The number of O VI absorption lines with $EW>100$km/s  is 
very small, while there are about $1-3$ lines per
unit redshift for O VII and O VIII absorption lines at $EW=100$km/s.
(3) Emission lines, primarily O VI and \lya in the UV and
O VII and O VIII in the soft X-rays 
are potentially observable by future missions and
they provide complimentary probes of the WHIM in 
often different domains of the temperature-density-metallicity 
phase space as well as in different spatial locations;
however, sometimes they overlap spatially and in phase space,
making joint analyses very useful.
The number of emission lines per unit redshift that may be detectable
by planned UV and soft X-ray missions are in the order of $0.1-1$.

\end{abstract}
 
 
\section{Introduction}

Cosmological hydrodynamic simulations have strongly suggested
that most of the previously ``missing" baryons may be in 
a gaseous phase in the temperature range
$10^5-10^7$K and at moderate overdensity 
(Cen \& Ostriker 1999, hereafter "CO"; Dav\'e 2001),
called the warm-hot intergalactic medium (WHIM),
with the primary heating process being
hydrodynamic shocks from the formation 
of large-scale structure at scales
currently becoming nonlinear.
The reality of the WHIM has now been quite convincingly
confirmed by a number of observations 
from HST, FUSE, Chandra and XMM-Newton
(Tripp, Savage, \& Jenkins 2000;
Tripp \& Savage 2000;
Oegerle \etal 2000;
Scharf \etal 2000;
Tittley \& Henriksen 2001;
Savage \etal 2002;
Fang \etal 2002; Nicastro \etal 2002;
Mathur, Weinberg, \& Chen 2002;
Kaastra \etal 2003;
Finoguenov, Briel, \& Henry 2003;
Sembach \etal 2004; Nicastro \etal 2005).

In addition to shock heating, feedback processes following
star formation in galaxies can heat gas to the same WHIM temperature range.
What is lacking theoretically is 
a satisfactory understanding of the known feedback processes
on the WHIM and how the WHIM may be used to 
understand and calibrate the feedback processes.
Another unsettled issue is how the predicted results
will change if one has a more accurate,
non-equilibrium calculation of the major
metal species, such as O VI, O VII and O VIII,
since time scales for ionization and recombination 
are not widely separated from the Hubble time scale.
In a companion paper (Cen \& Ostriker 2005) we have
studied the effects of galactic superwinds on the IGM,
and in particular on WHIM.
Additional improvements include
significantly larger dynamic ranges
of the simulation, a WMAP normalized cosmological model
and an improved radiative transfer treatment.
The purpose of this paper is to present
additional effects due to non-equilibrium calculations
on major observable metal species, such as O VI, O VII and O VIII.
Our current work significantly extends previous theoretical works
by our group and others 
(CO; Dav\'e 2001;
Cen \etal 2001; Fang \etal 2002; 
Chen \etal 2003;
Furlanetto \etal 2004,2005a,b;
Yoshikawa \etal 2003; Ohashi \etal 2004; Suto \etal 2004; Fang \et
al. 2005).
The outline of this paper is as follows:
the simulation details are given in \S 2;
in \S 3 we give detailed results
and discussion and conclusions are presented in \S 4.

\section{Simulations}\label{sec: sims}

The reader should refer to Cen \& Ostriker (2005)
for a detailed description of the computation method.
Here we will only briefly explain the basic hydrodynamic code,
explain the cosmological model that we use,
the simulation box parameters and 
relevant treatment details on how major oxygen species are computed.

The results reported on here are based on new simulations of
a {\it WMAP}-normalized (Spergel \etal 2003; Tegmark \etal 2004)
cold dark matter model with a cosmological constant, 
$\Omega_M=0.31$, $\Omega_b=0.048$, $\Omega_{\Lambda}=0.69$, $\sigma_8=0.89$,
$H_0=100 h {\rm km s}^{-1} {\rm Mpc}^{-1} = 69 {\rm km} s^{-1} {\rm Mpc}^{-1}$ 
and $n=0.97$.
The adopted box size is $85$Mpc/h comoving and with a number of cells
of $1024^3$, the cell size is $83$kpc/h comoving,
with dark matter particle mass equal to $3.9\times 10^8h^{-1}\msun$.
Given a lower bound of the temperature for almost
all the gas in the simulation ($T\sim 10^4$~K),
the Jeans mass $\sim 10^{10}\msun$ for mean density gas,
which is comfortably larger than our mass resolution.

As described in Cen \& Ostriker (2005), 
we have made simulations
with and without galactic superwinds (GSW).
Metals are produced self-consistently from star formation 
by adopting a specific efficiency of metal formation,
a ``yield" (Arnett 1996), $y_0=0.02$,
the percentage of stellar mass that is ejected back into IGM as metals
for Type II supernovae (SNe II),
which we follow accurately in the simulation.
On the other hand, accurately following the evolution of metal ejection
(approximately half of the iron)
from Type I supernovae (SNe I) is difficult in large part due to the uncertainties
in theoretical modeling of SNe I.
Since oxygen is predominantly produced by SNe II,
our results on oxygen related quantities are
very insensitive to the lack of treatment of SNe I.
Metals produced by SNe II are followed as a separate variable
(analogous to the total gas density) with the same hydrocode.
In addition, we implement another density variable to
keep track of the reprocessed, i.e., secondary metals in the ejecta
(such as S process elements),
which is proportional to the metallicity of the gas from 
which the star was formed.
There are two adjustable parameters with regard to
overall metal production, namely, $y_0$ from SNe II and $y_{I}$ from SNe I 
in terms of metal contribution.
These two parameters are uncertain theoretically but
may be normalized by observed oxygen and iron abundances in ICM.
The oxygen and iron abundance is 
$\sim 0.5$ and $\sim 0.3$ solar in the ICM of observed clusters
(e.g., Mushotzky \etal 1996; Tamura \etal 2004) and
there is indication that there is about an equal amount
of contribution to the iron mass in ICM from SNe Ia
and SNe II (e.g., Ettori 2005).
Following Tsujimoto \etal (1995), 
using metal yield patterns for
SNe Ia and SNe II with Salpeter IMF,
we find that in order to match the observed iron metallicity of $\sim 0.3$ solar in ICM,
we need a SNe II metal yield $y_0=0.03$ instead of 
the value $0.02$ adopted,
taking into account that one half of the iron in ICM would have been
contributed by SNe Ia.
However, this still leaves us with 
an oxygen abundance in the computed ICM of $\sim 0.25$ solar (using $y_0=0.03$)
versus the observed value of $\sim 0.5$ solar
(Mushotzky \etal 1996; Tamura \etal 2004).
This deficit in oxygen mass in ICM can not be made up by SNe Ia in any reasonable abundance,
because they do not produce a significant amount of oxygen mass.
We do not yet know the actual solution to this problem.
Possible solutions may include an IMF that is significant flatter than
the adopted Salpeter IMF, which would produce more SNe II
and more hypernovae, which are prime producers of oxygen.
Alternatively, one may increase the star formation efficiency by 
some factor, say, $\sim 2$,
to bring oxygen abundance into better agreement with observations,
which would then require some adjustment on the contribution to iron
mass in ICM from SNe Ia.
This is a very complex issue and there may be
hints that a simple combination of SNe II and SNe Ia 
with the standard IMF may not be adequate (e.g., Dupke \& White 2000;
Portinari \etal 2004; Baumgartner \etal 2005),
and it is beyond the scope of this paper to attempt
to address this issue.
What we must do, however, is to normalize the simulated abundances
of both iron and oxygen to the observed values in ICM as closely as we can.
This requirement translates to an enhancement of oxygen abundance
by a factor of $\sim 3=0.03/0.02\times 2$,
where the factor $0.03/0.02$ is needed to normalize to ICM iron abundance
(see Figure 15 below)
and the factor $2$ is to further normalize to ICM oxygen abundance;
we stress again that the underlying cause for the factor $2$ is unclear
at present.
Aside from this unknown scaling factor, oxygen is known to be predominantly
produced by SNe II and hence the computed spatial and temporal evolution 
of oxygen distribution should remain accurate;
we simply multiply the oxygen abundance at every spatial point in
the simulation by a factor of $3$.
Once this re-normalization procedure is done, 
we have no further freedom to make any other adjustment
with regard to metallicity, specifically, oxygen abundances,
in any other regions of the IGM, which are the focus of this study.
We have adopted a solar oxygen abundance of $8.78$ dex,
which is close to but about 20\% higher than
recently determined values (Allende Prieto, Lambert, \& Asplund 2001;
Asplund \etal 2004).

The implementation of the major oxygen species is done as follows.
We follow the evolution of O IV to O IX simultaneously
by directly integrating five rate equations for each cell
in a non-equilibrium fashion.
Each oxygen species is followed as a separate variable
and each conservation equation (without the source terms)
is solved using the same TVD hydrodynamics code,
taking into account
recombination (Shull \& van Steenberg 1982),
collisional ionization 
(Aldrovandi \& Pequignot 1973;
Shull \& van Steenberg 1982;
Arnaud \& Rothenflug 1985;
Arnaud \& Raymond 1992;
Verner \& Ferland 1996;
Voronov 1997)
and photoionization (Clark, Cowan, \& Bobrowicz 1986;
Verner \etal 1996)
processes for a non-uniform cosmological density field.
Relevant coefficients for these processes and equations 
are enclosed in the Appendix for completeness.

\section{Results}

We present and compare simulations with and without
galactic superwinds (GSW; Cen \& Ostriker 2005) and
between calculations that assume ionization-equilibrium
and detailed non-equilibrium calculations.

\subsection{Metal Absorption Lines in UV and X-ray}

We present detailed results on major absorption lines,
including O VI ($\lambda \lambda$1032, 1038), 
O VII ($\lambda $21.6) and O VIII ($\lambda $19.0).
The detailed evolution of all these major metal species is followed
in a non-equilibrium fashion, by solving rate equations for each cell at each timestep
explicitly. The evolution also depends on the detailed
background radiation field as a function of redshift,
which we compute in a self-consistent way in our simulations
taking into account both sources and sink computed directly in simulations,
shown in Figure 1.
All our results presented below are at $z\sim 0$ and hence
the z=0 radiation field is used when we compute metal abundances
for some species based on CLOUDY (equilibrium) method.
We note that the computed $z=0$ radiation field is
consistent with observations (Shull \etal 1999) and so is
the radiation field at higher redshifts (Rauch \etal 1997; Bolton \etal 2005)
and with observationally derived evolution of 
the background radiation field (Haardt \& Madau 1999).

Figure 2 shows the number of O VI absorption lines per unit redshift
as a function of equivalent width from simulations
with and without GSW, from the case based on the simulation 
using CLOUDY program based on the simulation with GSW,
all compared to observations (symbols).
First, we note that the simulation with appropriate
GSW and non-equilibrium calculation (solid curve)
is in excellent agreement with direct observations (symbols).
This is quite remarkable considering that we have not attempted
to make any fine-tuning on model parameters and
the only two major parameters that have significant
effects are $e_{GSW}$ and $y$, where
the latter is normalized to observed X-ray cluster gas abundance
and the former is chosen to match the observed galactic superwinds
(e.g., Pettini \etal 2002).
In other words, our model is basically free of adjustable
parameters other than overall yield and $e_{GSW}$
and it is fair to say that the excellent agreement between our calculation and 
observations is a strong indication that our cosmological
model is a faithful representation of the true universe in this regard.

However, the excellent agreement would not have been found,
had the GSW effect not been included (dotted curve)
or if we had adopted the equilibrium (CLOUDY) treatment is performed (dashed curve).
Both GSW feedback and non-equilibrium method produce 
significant errors for abundances of low equivalent widths absorption lines 
($EW\le 30\AA$).
Fortuitously, one may obtain an apparently reasonable agreement between 
theoretical calculations and observations
if both errors in modeling are simultaneously made, i.e.,
an equilibrium CLOUDY treatment on a simulation with no significant GSW feedback;
this is a mere coincidence.
The physical reason why the equilibrium treatment 
produced more O VI than non-equilibrium simulation
is that collisional ionization time scales are longer
than the time scale on which gas is shock heated,
especially for low density regions;
for the very high density regions, the two approaches should agree,
as Figure 2 indicates.
Results obtained here are broadly consistent with 
with earlier simulations (Cen \etal 2001; Fang \& Bryan 2001; Chen \etal 2003)
and with analytic work (Furlanetto \etal 2005b)
with regard to O VI absorption line abundance,
taking into account various uncertainties 
about cosmological model parameters,
different treatments of metallicity and assumptions
about ionization equilibrium.

Figure 3 shows the number of O VII absorption lines per unit redshift
as a function of equivalent width from simulations
with (solid curve) and without (dotted curve) GSW.
Also included as a dashed curve is what one would obtain
using CLOUDY program based on the simulation with GSW.
Somewhat different from the case for O VI,
we see here that the feedback and non-equilibrium effects
each produce a $30-50\%$ change in the opposite direction
in the predicted number of O VII absorption
lines for the entire range of width.
Figure 4 is similar to Figure 3 but
for O VIII absorption lines and comparable effects due to
GSW and equilibrium treatment are seen.
In both cases for O VII and O VIII absorption lines
the agreement between observations and our detailed non-equilibrium calculations with GSW
are not as good as that for the O VI absorption line see in Figure 2.
However, the observational errorbars for the abundances of these two absorption lines
are currently very large due to a very small observational sample.
The physical reason behind the results that the equilibrium treatment 
produced more O VII and O VIII than non-equilibrium simulation
is similar to that given for the case of O VI,
except that the collisional ionization time scales are still longer
in the case of O VII and O VIII and the times scales on which gas is shock heated
are still shorter than for O VI.
as Figure 2 indicates.
It will be of great importance to enlarge observational samples
for both O VII and O VIII absorption lines to make more statistically 
sound comparisons.
Since our model has no adjustable parameters except
yield and $e_{GSW}$, 
the outcome from such
comparisons will provide critical tests of 
yield, $e_{GSW}$ and the underlying cosmological model.

To summarize the relative abundances of 
these major species,
Figure 5 displays results for all three absorption lines
in terms of equivalent widths in units of $km/s$ to indicate
the relative abundances of the three major species.
The conversion between EW in m\AA\ and EW in km/s is:
$EW(m\AA)= \lambda~EW(km/s)/c$,
which is equal to $(34.3m\AA)(EW(km/s)/10km/s)$,
$(0.35m\AA)(EW(km/s)/50km/s)$
and $(0.32m\AA)(EW(km/s)/50km/s)$,
for O VI, O VII and O VIII, respectively.
The expected abundances for these three lines are in the range
$50-100$ per unit redshift at $EW=1$km/s decreasing to 
$10-20$ per unit redshift at $EW=10$km/s; the rate of decrease
of the line abundance at higher $EW$ then becomes significantly higher.
The number of O VI absorption lines with $EW>100$km/s  is 
very small, while there are about $1-3$ lines per
unit redshift for O VII and O VIII absorption lines at $EW=100$km/s.
The computed O VII and O VIII absorption line
abundances are in reasonable agreement with 
Chen \etal (2003), in particular, 
for their ``scatter" metallicity model (dashed curve in Figure 7 of Chen 
\etal 2003), considering large uncertainties on 
metallicity distribution and ionization equilibrium assumption.

Figure 6 
shows the average metallicity (with scatter) of O VI, O VI and O VIII absorption 
lines as a function of equivalent widths.
We see two clear trends, albeit with significant scatter:
(1) at low EW on average O VI absorption lines
have lowest metallicity, while  
O VIII absorption lines have highest metallicity.
(2) for each absorption line higher EW lines have higher metallicity.
Overall, on average, O VI lines have a metallicity
range of $0.15-0.60\zsun$ with a large scatter at the low end.
The O VII absorption lines 
have a mean metallicity range 
$0.20-0.60\zsun$ with a somewhat larger scatter at the low end 
than the O VI line.
The O VIII absorption lines, quite interestingly,
display a nearly constant mean metallicity 
of $\sim 0.5-0.6\zsun$ over the entire EW range.
All three absorption lines converge to a mean metallicity 
of $0.6\zsun$ at the high EW end with a scatter of about $0.2\zsun$.
These figures, (19)-(24),
present one of the two primary observational predictions of this paper.

Figure 7 shows the Doppler width $b$ as a function of equivalent 
width $EW$ for O VI, O VII and O VIII, from top to bottom panel.
We see that the Doppler widths for the three species 
all have a concentration broadly in  
the range of $10-60$~km/s at the low $EW$ end.
This suggests that peculiar velocity effect plays a very important role here.
It should be noted that peculiar velocities affects
widths in complex ways, either broadening or narrowing,
depending on detailed velocity and shock structures.
In some cases, Hubble expansion could counter the effect 
of peculiar velocities of enclosing double shocks.
Most of the very low $b$ structures are likely photo-ionized
and overall velocity broadening effect is minimal.
On the high $EW$ end, the trend is somewhat more clear.
There, values of $b$ of O VI, O VII and O VIII 
appear to increase more or less monotonically in that order.
This is due to the fact that higher ionization species
are closer to center of potential wells, thus
have, on average, higher temperatures and velocities.

Finally, Figures 8,9,10 show in detail three randomly chosen
lines of sight 
through the simulation boxes
that have significant absorption features.
Noticeable features in these figures are:
(1) GSW in most cases tend to broaden the absorption line profiles.
(2) GSW may alter absorption profile in complex ways as clearly
seen in the bottom two panels of Figure 8 for O VII and O VIII lines,
caused by a combination of separate effects on the gas velocity,
temperature, metallicity and density.
(3) The typical gas density of the absorption features seen
is in the range of $10-300$ times the mean gas density.
(4) The typical metallicity of gas producing the prominent
UV absorption lines is in the range $0.1-1.0\zsun$.
(5) The significant absorption features are usually
enclosed by double shocks with velocity in the range
$100-400$km/s,
i.e., we are seeing the outer edges of a cooling ``Zeldovich pancake".
In fact, such pattern has been detected in a
cluster of multiphase absorbers along the sightline toward
PKS~2155-304 (Shull \et al. 2003).
We note that multiphase medium, while common, most often does not
spatially co-exist; rather, complex velocity, density, and metallicity
structure mask their appearances when observed in Hubble space.
(6) The Doppler line widths for deep lines have $100-300$km/s,
while for more abundant small lines the widths are small,
in the range of $1-100$km/s.
(7) The relative absorption strengths of the three considered lines
vary and no particular order is visible, as expected,
since the line strengths are functions of several physical variables.
More detailed studies will be defered to a subsequent paper.

\subsection{Emission Lines in UV and X-ray}

In order to map out directly the WHIM structure in 3 dimensions,
detailed emission measurements are invaluable, 
because of their potential for
continuous coverage and a lack of a dependence on the sparsely
distributed background sources as required for absorption studies.
We will focus on three metal emission lines, O VI, O VII and O VIII as well
as the \lya emission line.
Emission tables were generated using 
a software package called CHIANTI (http://wwwsolar.nrl.navy.mil/chianti.html),
based on electron density and density of a specific metal species.

Figure 11 shows the cumulative number of O VI and \lya lines in UV
per unit redshift as a function of surface brightness.
The red curves show results from the simulation
with GSW and the green curve from the simulation without GSW.
The black curve (for O VI line only) is computed using CLOUDY code
on the assumption of ionization equilibrium
based on the density, temperature
and metallicity information from the simulation with GSW.
The results found with regard to \lya emission may be compared to
Figure 8 of Furlanetto \etal (2005a). Their figure has a different
units for the ordinate, while the abscissa has the same units.
In order to convert the values 
of the ordinates in their Figure 8 to $dn/dz$ displayed here,
we need to multiply their value in the vertical axis by 
$(dl/dz)/\delta l =4478$, where $\delta l=0.67h^{-1}$Mpc 
is their pixel depth and $dl/dz=3000h^{-1}$Mpc is the length
per unit redshift at $z=0$.
Good agreements are found between their results and ours.
For example, at surface 
brightness of $\phi=10^2$ photons~cm$^{-2}$~s$^{-1}$~sr$^{-1}$,
we find $dn/dz(>\phi)=0.33$ (thin solid curve in Figure 11), 
while their values range
from $0.11$ (panel b of their Figure 8) to $0.70$ (panels a and c
of their Figure 8).
Also, we may compare the results found with regard to O VI emission 
to Figure 9 of Furlanetto \etal (2004), which shows the PDF
(defined as $dn/d\ln\phi$ for $\delta z=0.01$)
of O VI emission.
Integrating, for example, the dotted curve
in panel (c) of their Figure 9, we find their
$dn/dz (>\phi)=0.40$ at surface 
brightness of $\phi=10^2$ photons~cm$^{-2}$~s$^{-1}$~sr$^{-1}$,
which should be compared to our value of $0.61$ (thick solid curve in Figure 11).
At face value, their emission estimate is somewhat lower than ours
but the difference should be considered to be small,
given many physical and numerical factors involved.
Therefore, we instead consider this a good agreement.

Figure 12 shows a similar plot for O VII and O VIII lines in soft X-rays.
Note that future planned UV missions (for O VI and \lya lines)
may be able
to achieve a sensitivity in the range $100-1000$ in the displayed units
and the proposed soft X-ray missions (for O VII and O VIII lines;
DIOS -- Yoshikawa \etal 2003; MBE -- Fang \etal 2005)
is expected to be able
to achieve a sensitivity of order $0.1$ in the displayed units for
a reasonable amount of exposure time.
We see that both UV emission lines such as \lya, O VI and
soft X-ray emission lines such as O VII and O VIII
have comparable abundance.
The number of emission lines per unit redshift that may be detectable
by planned UV and soft X-ray missions 
are in the order of $0.1-1$.
Therefore, it is highly desirable to be able to 
map out emission in both UV and X-ray bands,
because they often probe complimentary physical regions.

How large a fraction of the baryons may be probed by these
metal emission lines?
Figures 13,14 show
the total amount of baryons that can be probed by 
emission lines as a function of the surface brightness.
For the quoted sensitivities in the range $100-1000$ in the displayed units,
we see that O VI line 
may be able to probe about $0.5-1.5\%$ of all baryons
or $1-3\%$ of WHIM, while the \lya emission line
may be able to probe about $4-6\%$ of WHIM.
The DIOS and MBE X-ray mission is expected to be able
to detect $10\%$ and $20\%$ of WHIM, using 
O VII and O VIII lines, respectively,
which may be compared to calculations
by Yoshikawa \etal (2004b) who
found that at $10^{-10}$ erg~s$^{-1}$~cm$^{-2}$~sr$^{-1}$
(which corresponds to our adopted DIOS and MBE sensitivity of
$\phi=0.1$ photons~cm$^{-2}$~s$^{-1}$~sr$^{-1}$)
the WHIM mass fractions probed O VII and O VIII
are $18\%$ and $20\%$, respectively.
Given the difference in the treatments between ours and theirs
the agreement is to be considered good
and this also suggest that the local universe, as they simulated,
represents a fair sample of the universe with regard to WHIM.
Since metal enrichment of the IGM is quite inhomogeneous,
these emission lines, being in overdense regions,
should be able to track a larger fraction of metals produced
in star formation.
Figures 15,16 show that 
the O VI emission line
will be able to probe about $1.5-4\%$ of all metals in the IGM, 
while the O VII and O VIII emission lines
will be able to probe about $15-30\%$ of all metals in the IGM.
In all the above figures (Figures 11-16) we show that 
GSW affect quantitative results dramatically; omission of GSW effect
could cause errors $100-1000\%$ with regard to line abundances,
baryonic densities and metal densities.

Figures 17,18 show the observable regions in the sky (black regions)
of a map of size $85\times 85$Mpc$^2$/h$^2$ 
using O VI and  \lya lines, respectively,
with an instrument of a sensitivity of 100~photon~cm$^{-2}$sr$^{-1}$s$^{-1}$.
Figures 19,20 show the observable regions in the sky (black regions)
of a map of size $85\times 85$Mpc$^2$/h$^2$ 
using O VII and O VIII lines, respectively,
with an instrument of a sensitivity of 0.1~photon~cm$^{-2}$sr$^{-1}$s$^{-1}$
(DIOS).
The red contours are the underlying gas density distribution.
It is quite clear that, while detectable regions by these UV and soft 
X-ray emission lines can not continuously cover the entire filamentary network,
they provide a faithful representation of the underlying 
density structure, especially the filaments.
It may be noted that O VI and \lya lines provide 
rather poor tracer for some the 
of the larger red contour concentrations, because the latter
are sites of hotter gas in the WHIM temperature range,
too hot to be traceable by
these UV emission lines; O VII and O VIII X-ray emission lines
provide a relatively better set of probers for this hot gas.
Figures 21,22,23,24 zoom in to a smaller region of Figure 17,18,19,20
respectively, to better display the detailed correspondence
between detectable regions and the underlying density structure.
We see that both O VI and \lya lines tend to avoid the larger density knots
and often appear to be off-centered, i.e., located in the outskirts of large density
concentrations.
In contrast, 
the O VII and  O VIII emission lines most often directly overlap with
the larger density concentrations and do not probe the outskirts of them.
This clearly illustrates the complimentarity of UV and X-ray emission missions,
dictated by the complex physical structure of the WHIM in the vicinity
of galaxies and groups of galaxies.
The physical variables that play important roles here are gas temperature,
gas density, gas metallicity and gas peculiar velocity.
None of these can be tracked simply but all of them 
are expected to be influenced significantly by GSW.
Therefore, detailed comparisons between simulations and observations,
when becoming available,
may provide an extremely valuable tool to probe galaxy formation
and its all important feedback processes.

\section{Discussion and Conclusions}

In Cen \& Ostriker (2005)  
we show that our significantly improved simulations confirm 
previous conclusions based on earlier simulations:
{\it nearly one half of all baryons at the present epoch
should be found in the WHIM} - a filamentary network 
in the temperature range of $10^5-10^7$~K.
There, we also presented significant effects due to feedback from star formation.

Here we investigate detailed ionization distributions of 
oxygen, computed explicitly in a non-Equilibrium way,
and present results related to 
important UV and soft X-ray lines and how their observations
may shed light on WHIM and galaxy formation processes.
Our finding are:
(1) non-equilibrium calculations produce
significantly different results from equilibrium calculations, and
the differences are complex and difficult to characterize simply.
(2) The abundance of O VI absorption lines based on
non-equilibrium simulations with galactic superwinds 
is in remarkably good agreement with latest observations,
implying the validity of our model,
while the predicted abundances for O VII and O VIII absorption lines
appear to be lower than observed but the observational errorbars
are currently very large.
The expected abundances for O VI (as well as \lya), O VII and O VIII lines 
are in the range $50-100$ per unit redshift at $EW=1$km/s decreasing to 
$10-20$ per unit redshift at $EW=10$km/s.
The number of O VI absorption lines with $EW>100$km/s  is 
very small, while there are about $1-3$ lines per
unit redshift for O VII and O VIII absorption lines at $EW=100$km/s.
(3) Emission lines, primarily O VI and \lya in the UV and
O VII and O VIII in the soft X-rays 
are potentially observable by future missions and
they provide complimentary probes of the WHIM in 
often different domains in the temperature-density-metallicity 
phase space as well as in different spatial locations;
however, sometimes they overlap spatially and in phase space,
making joint analyses very useful.
The number of emission lines per unit redshift that may be detectable
by planned UV and soft X-ray missions are in the order of $0.1-1$.

We expect that future missions in UV and X-ray 
should be able to provide much more accurate characterization
of the WHIM through two complimentary approaches:
major UV and soft X-ray absorption lines and
emission lines.
The former will be able to trace out bulk of the mass 
as well as volume occupied by the WHIM,
whereas the latter will be able to probe higher density regions.
Together they may allow us to construct a coherent
picture of the evolution of the IGM.
In addition, detailed useful information may be obtained
by investigating relations between galaxies and
properties of the WHIM, keeping in mind that
GSW play an essential role exchanging 
mass, metals and energy between them.
We stress that proper comparisons between observations and simulations
in the vicinity of galaxies where GSW effects originate and are strongest
may provide one of the most useful ways to probe star formation
and its all important feedback processes.

\acknowledgments

We thank Ed Jenkins for kindly providing
emission tables for oxygen species.
We thank Ed Jenkins, Jerry Ostriker and Mike Shull for useful comments.
The simulations were performed at
the Pittsburgh Supercomputer Center.
We thank R. Reddy at 
the Pittsburgh Supercomputer Center
for constant and helpful assistance 
in the process of making the simulations.
This work is suppoted in part by grants NNG05GK10G
and AST-0507521. T.~Fang was supported by the NASA through
{\sl Chandra} Postdoctoral Fellowship Award Number PF3-40030 issued by the
{\sl Chandra} X-ray Observatory Center, which is operated by the
Smithsonian Astrophysical Observatory for and on behalf of the NASA under
contract NAS8-39073.

\newcommand{\so}{\mbox{\scriptsize{$\rm O$\ }}}
\newcommand{\sov}{\mbox{\scriptsize{$\rm O~V$\ }}}
\newcommand{\sovi}{\mbox{\scriptsize{$\rm O~VI$\ }}}
\newcommand{\sovii}{\mbox{\scriptsize{$\rm O~VII$\ }}}
\newcommand{\soviii}{\mbox{\scriptsize{$\rm O~VIII$\ }}}
\newcommand{\sovix}{\mbox{\scriptsize{$\rm O~IX$\ }}}
\begin{appendix}

\section{Equations for Non-Equilibrium Evolution of Oxygen Species}

The ionization fractions of major oxygen species are calculated in a
non-equilibrium evolution fashion. Specifically, we calculate the
ionization fractions from \ion{O}{5} to \ion{O}{9} by considering
three major processes: (1) photoionization; (2) collisional
ionization; and (3) recombination. 
If we define the ionization
fraction of species $i$ is $X_i$, the time-depended evolution of $X_i$
is determined by
\begin{equation}
\frac{dX_i}{dt} = -X_i \Gamma_i - \alpha_i X_i n_e - \beta_i X_i n_e +
X_{i-1} \Gamma_{i-1} + \alpha_{i+1} X_{i+1} n_e + \beta_{i-1} X_{i-1} n_e.
\end{equation}
\noindent 
Here $\Gamma$ is the photoionization rate of, given by
\begin{equation}
\Gamma = \int_{\nu_0}^{\infty} \frac{4 \pi J_{\nu}}{h\nu} \sigma(\nu) d\nu,
\end{equation}
\noindent 
where $\nu_0$ is the ionization frequency, $J_{\nu}$ is the background
radiation, and $\sigma$ is the photoionization cross section, which
we adopted from Verner et al.~(1996). Here $n_e$ is the electron
density, $\alpha$ is the recombination rate adopted from Shull \&
Steenberg~(1982), and $\beta$ is the collisional ionization rate
adopted from Voronov~(1997). 
We solve the combined set of equations for species
\ion{O}{1} to \ion{O}{4} using a first-order implicit
scheme, necessary for the stiff equations considered here.
Given a time step of $\Delta t$,
the ion number density at step $N+1$ are determined by the
following set of equations:

\begin{equation}
\begin{array}{llll}
n_{\sov}^{N+1} &=& n_{\sov}^{N} &-\ (n_{\sov}^{N+1}n_e^N\beta_{\sov}+n_{\sov}^{N+1}
\Gamma_{\sov})\Delta t \\
& & &+\ n_{\sovi}^{N+1}n_e^N\alpha_{\sov}\Delta t \\
\\
n_{\sovi}^{N+1} &=& n_{\sovi}^{N} &-\
(n_{\sovi}^{N+1}n_e^N\beta_{\sovi}+n_{\sovi}^{N+1}n_e^N\alpha_{\sov}+n_{\sovi}^{N+1}\Gamma_{\sovi})\Delta
t\\
& &
&+\ (n_{\sov}^{N+1}n_e^N\beta_{\sov}+n_{\sovii}^{N+1}n_e^N\alpha_{\sov}+n_{\sov}^{N+1}\Gamma_{\sov})\Delta
t\\
\\
n_{\sovii}^{N+1} &=& n_{\sovii}^{N} &-\
(n_{\sovii}^{N+1}n_e^N\beta_{\sovii}+n_{\sovii}^{N+1}n_e^N\alpha_{\sovi}+n_{\sovii}^{N+1}\Gamma_{\sovii})\Delta
t\\
& &
&+\ (n_{\sovi}^{N+1}n_e^N\beta_{\sovi}+n_{\soviii}^{N+1}n_e^N\alpha_{\sovi}+n_{\sovi}^{N+1}\Gamma_{\sovi})\Delta
t\\
\\
n_{\soviii}^{N+1} &=& n_{\soviii}^{N} &-\
(n_{\soviii}^{N+1}n_e^N\beta_{\soviii}+n_{\soviii}^{N+1}n_e^N\alpha_{\sovii}+n_{\soviii}^{N+1}\Gamma_{\soviii})\Delta
t\\
& &
&+\ (n_{\sovii}^{N+1}n_e^N\beta_{\sovii}+n_{\sovix}^{N+1}n_e^N\alpha_{\sovii}+n_{\sovii}^{N+1}\Gamma_{\sovii})\Delta
t\\
\end{array}
\end{equation}
\noindent 
the above set of the equations is completed by 
\begin{equation}
n_{\sovix} = n_O - (n_{\sov}+n_{\sovi}+n_{\sovii}+n_{\soviii}).
\end{equation}
\noindent 
To simplify notations, we define
\begin{equation}
x \equiv n_{\sov}; y \equiv n_{\sovi}; z \equiv n_{\sovii}; w \equiv n_{\soviii},
\end{equation} 
\noindent 
and the final solution to Eq.(A3) is then
\begin{equation}
\begin{array}{lll}
x^{N+1} &=& b^7x^N + b^8y^N + b^9z^N + b^{10}w^N +
b^{10}a^{11}n_{\so}\\ y^{N+1} &=& \left(b^1b^7 - \frac{1}{a^2}\right)
+ b^1b^8y^N + b^1b^9z^N + b^1b^{10}w^N+ b^1b^{10}a^{11}n_{\so}\\
z^{N+1} &=& \left(b^2b^7 - b^3\right) + \left(b^2b^8 -
\frac{1}{a^5}\right)y^N + b^2b^9z^N + b^2b^{10}w^N+
b^2b^{10}a^{11}n_{\so}\\
w^{N+1} &=& \left(b^4b^7 + b^5\right) + \left(b^4b^8 -
b^6\right)y^N + \left(b^4b^9 -
\frac{1}{a^8}\right)z^N + b^4b^{10}w^N+
b^4b^{10}a^{11}n_{\so}\\
\end{array}
\end{equation}

Here we define:
\begin{equation}
\begin{array}{lllllllll}
a^1 &\equiv& n_e^N \beta_{\sov} \Delta t; & a^2 & \equiv & n_e^N
\alpha_{\sov} \Delta t; & a^3 & \equiv & \Gamma_{\sov} \Delta t\\
a^4 &\equiv& n_e^N \beta_{\sovi} \Delta t; & a^5 & \equiv & n_e^N
\alpha_{\sovi} \Delta t; & a^6 & \equiv & \Gamma_{\sovi} \Delta t\\
a^7 &\equiv& n_e^N \beta_{\sovii} \Delta t; & a^8 & \equiv & n_e^N
\alpha_{\sovii} \Delta t; & a^9 & \equiv & \Gamma_{\sovii} \Delta t\\
a^{10} &\equiv& n_e^N \beta_{\soviii} \Delta t; & a^{11} & \equiv & n_e^N
\alpha_{\soviii} \Delta t; & a^{12} & \equiv & \Gamma_{\soviii} \Delta t\\
\end{array}
\end{equation} and
\begin{equation}
\begin{array}{lcl}
b^1 & \equiv & (1+a^1+a^3)(a^2)^{-1} \\
b^2 & \equiv & \left[(1+a^1+a^3)(1+a^2+a^4+a^6)-(a^1+a^3)a^2\right](a^2a^5)^{-1} \\
b^3 & \equiv & (1+a^2+a^4+a^6)(a^2a^5)^{-1} \\
b^4 & \equiv & \left[(1+a^5+a^7+a^9)b^2-(a^4+a^6)b^1\right](a^8)^{-1}\\
b^5 & \equiv & \left[(a^4+a^6)-(1+a^5+a^7+a^9)a^2b^3\right](a^2a^8)^{-1}\\
b^6 & \equiv & (1+a^5+a^7+a^9)(a^5a^8)\\
b^7 & \equiv & \left[a^{11}/a^2 - (a^7+a^9-a^{11})b^3 - (1+a^8+a^{10}+a^{11}+a^{12})b^5\right]\\
        &             & \left[(1+a^8+a^{10}+a^{11}+a^{12})b^4+a^{11}+a^{11}b^1-(a^7+a^9-a^{11})b^2\right]^{-1}\\
b^8 & \equiv & \left[(1+a^8+a^{10}+a^{11}+a^{12})b^6 - (a^7+a^9-a^{11})/a^5\right]\\
        &             & \left[(1+a^8+a^{10}+a^{11}+a^{12})b^4 + a^{11}+a^{11}b^1-(a^7+a^9-a^{11})b^2\right]^{-1}\\
b^9 & \equiv & \left[(1+a^8+a^{10}+a^{11}+a^{12})/a^8\right]\\
        &             & \left[(1+a^8+a^{10}+a^{11}+a^{12})b^4 + a^{11}+a^{11}b^1-(a^7+a^9-a^{11})b^2\right]^{-1}\\
b^{10} & \equiv & \left[(1+a^8+a^{10}+a^{11}+a^{12})b^4 + a^{11}+a^{11}b^1-(a^7+a^9-a^{11})b^2\right]^{-1}\\                
\end{array}
\end{equation}

We test our equations by 
comparing calculations against ionization fractions obtained from
various literatures and/or output from CLOUDY (Ferland et al.~1998),
based on three cases: (1) collisional ionization only; 
(2) including additional photonization
with a background radiation field of $J(912\AA) = 10^{-22} \rm\ ergs\
s^{-1}cm^{-2}Hz^{-1}sr^{-1}$; 
and (3) photonization with a background
radiation field of $J(912\AA) = 10^{-23} \rm\ ergs\
s^{-1}cm^{-2}Hz^{-1}sr^{-1}$. 
Here $J(912\AA)$ is the specific flux at
912 $\AA$,\ and we adopt a power law spectrum with a spectral index of $-1$. 
To make the comparison to available equilibrium calculations,
we run our simulations long enough so
eventually all the ion species are in ionization equilibrium. 
Figure~25 shows the comparison for the collisional ionization case
with zero background radiation field. 
The solid lines in each
panel are results from our calculation, and the diamonds are adopted
from Mazzotta et al.~(1998). Except \ion{O}{5}, all the other ion
species show excellent agreements between our calculations and those from
Mazzotta et al.~(1998), which are calculated based on collisional
ionization equilibrium. 
Figures 26 and 27 show the comparison for case
(2) and (3), respectively. The symbols are the same as those in
Figure~25. These comparisons clearly demonstrate that our simulation
reproduces the ionization fractions under a variety of possible
cases that have bearing on our actual calculations.

\end{appendix}

\begin{figure}
\plotone{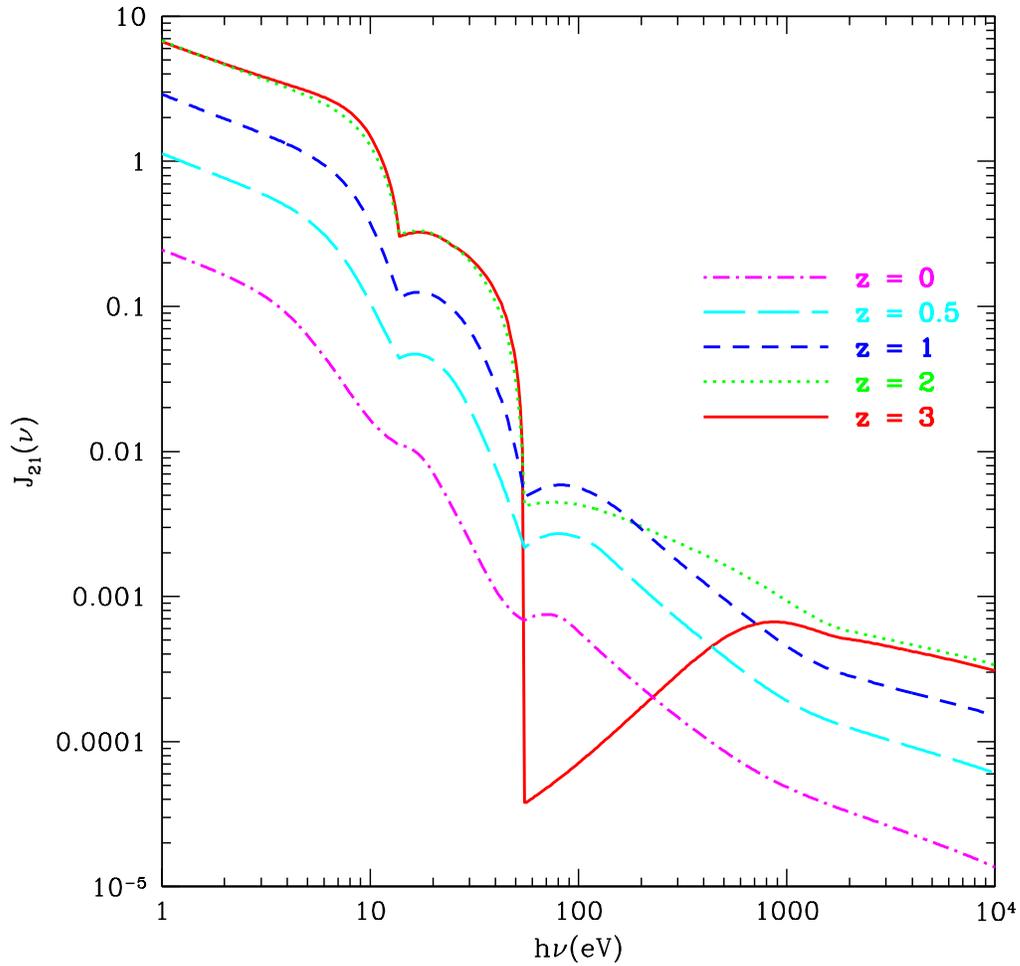}
\caption{
shows the background radiation field
in units of $10^{-21}$erg/cm$^2$/sec/sr,
at five redshift, z=3,2,1,0.5,0, computed in our simulations.
}
\label{f5}
\end{figure}

\begin{figure}
\includegraphics[angle=90.0,scale=0.65]{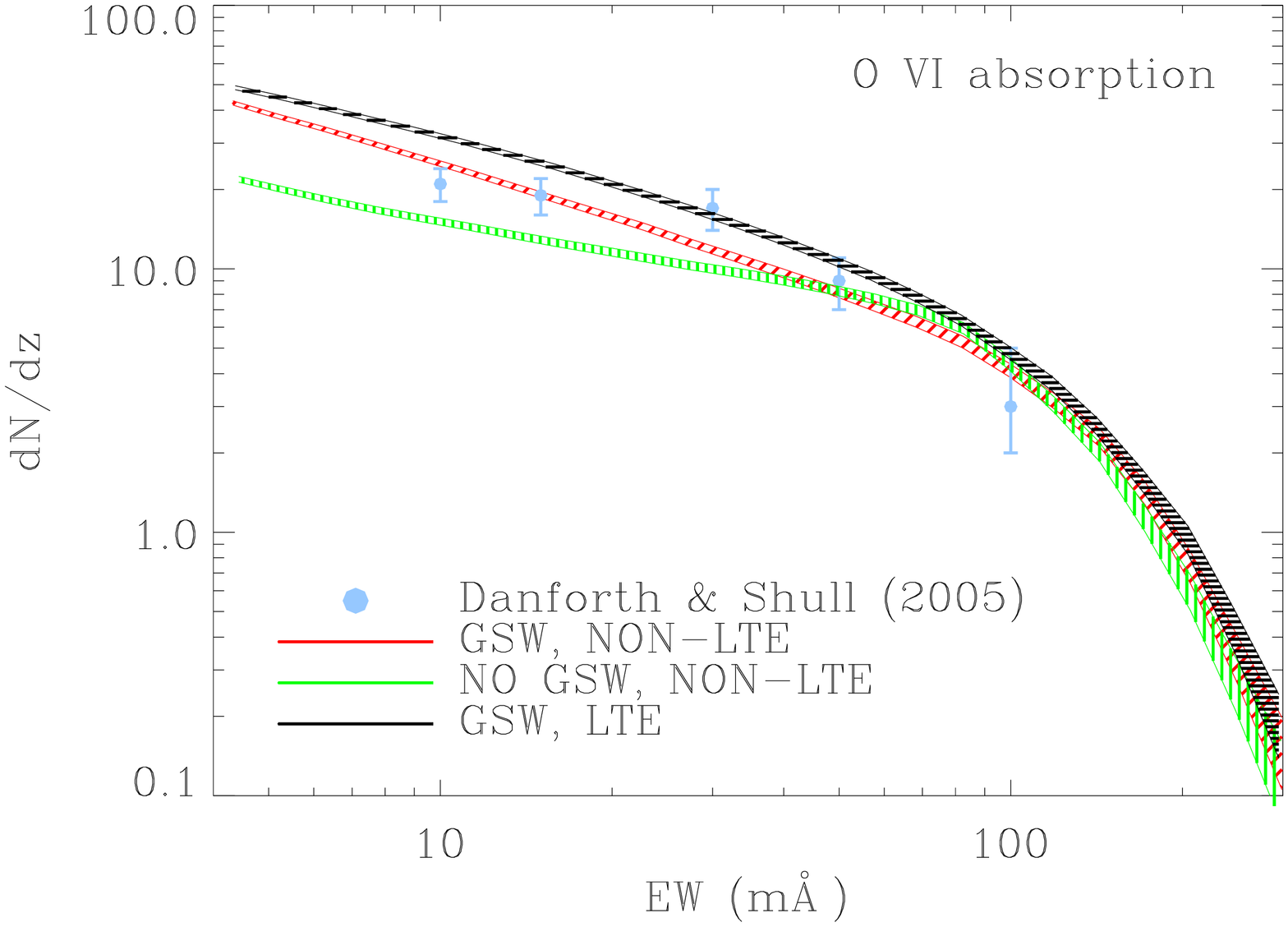}
\caption{
shows the number of O VI absorption lines per unit redshift
as a function of equivalent width in units of m\AA.
The red curve shows our primary results from the simulation
with GSW and the green curve from the simulation without GSW.
The black curve is computed using CLOUDY code
on the assumption of ionization equilibrium
based on the density, temperature
and metallicity information from the simulation with GSW.
The symbols are observations by Danforth \& Shull (2005).
}
\label{f5}
\end{figure}

\begin{figure}
\includegraphics[angle=90.0,scale=0.65]{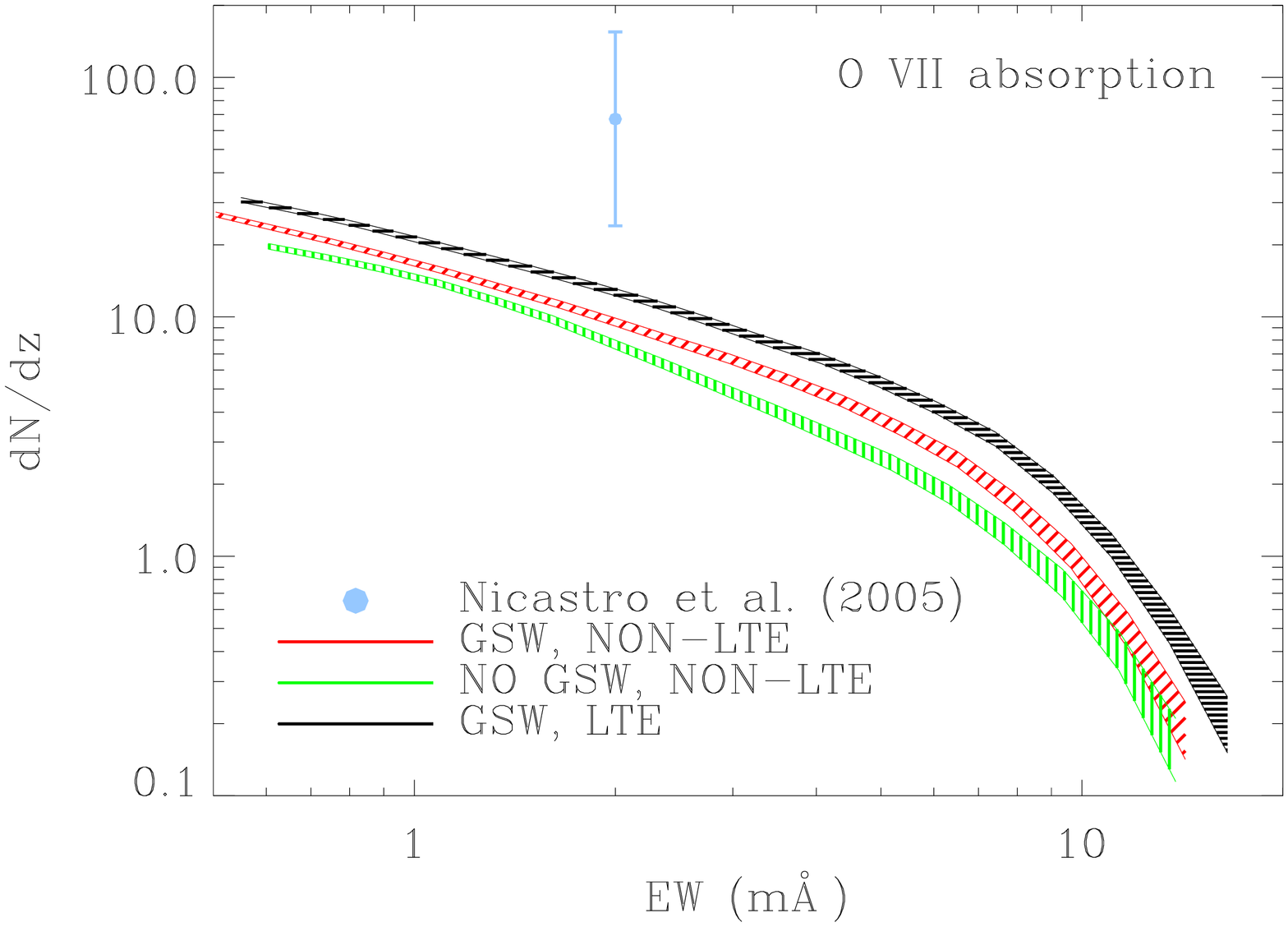}
\caption{
shows the number of O VII absorption lines per unit redshift
as a function of equivalent width in units of m\AA.
The red curve shows results from the simulation
with GSW and the green curve from the simulation without GSW.
The black curve is computed using CLOUDY code
on the assumption of ionization equilibrium
based on the density, temperature
and metallicity information from the simulation with GSW.
The symbols are observations by Nicastro \etal (2005).
}
\label{f6}
\end{figure}

\begin{figure}
\includegraphics[angle=90.0,scale=0.65]{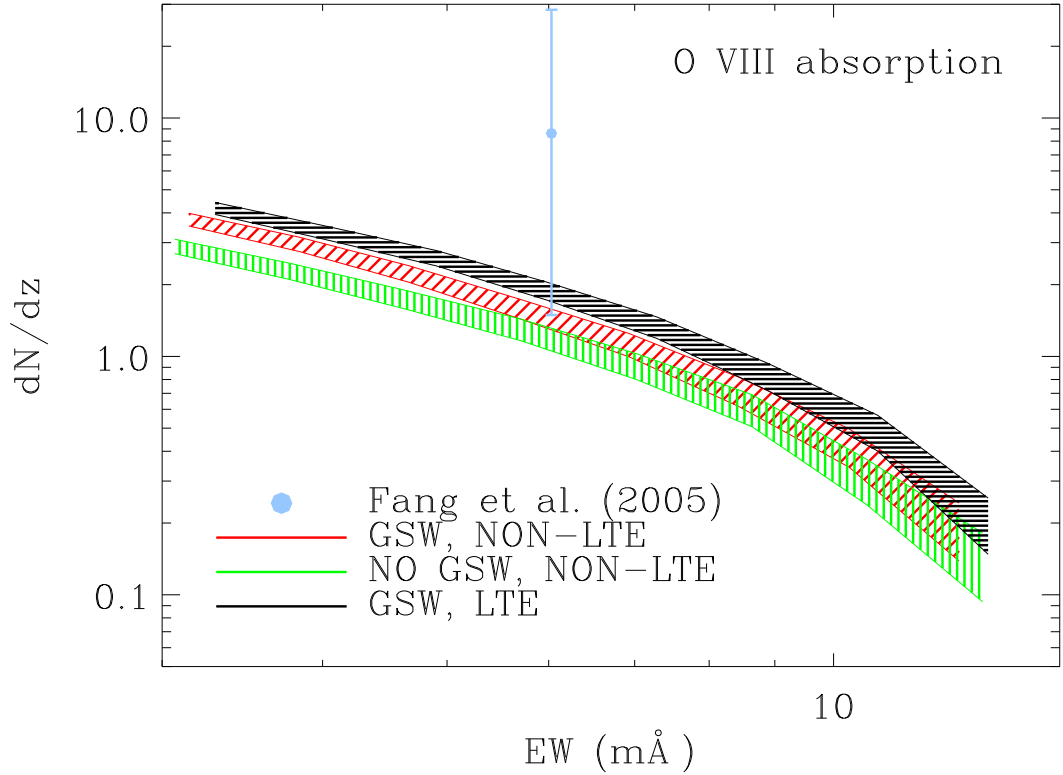}
\caption{
shows the number of O VIII lines per unit redshift
as a function of equivalent width.
The red curve shows results from the simulation
with GSW and the green curve from the simulation without GSW.
The black curve is computed using CLOUDY code
on the assumption of ionization equilibrium
based on the density, temperature
and metallicity information from the simulation with GSW.
The symbols are observations by Fang \etal (2005).
}
\label{f7}
\end{figure}

\begin{figure}
\includegraphics[angle=90.0,scale=0.65]{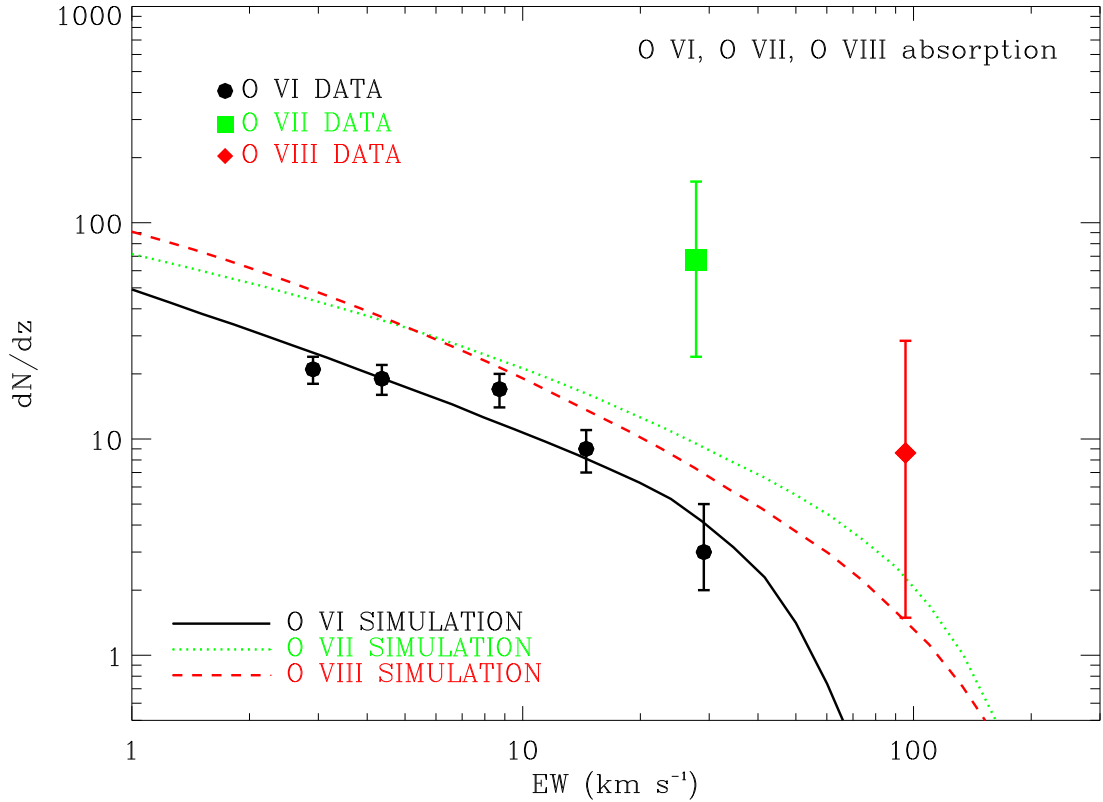}
\caption{
shows the number of all three absorption lines, O VI (solid), O VII (dotted) and O VIII (dashed), 
per unit redshift as a function of equivalent width
in units of km/s, from the simulation with GSW.
The symbols are observations by Danforth \& Shull (2005),Nicastro \etal (2005)
and Fang \etal (2005)
for O VI, O VII and O VIII absorption lines, respectively.
}
\label{f8}
\end{figure}

\begin{figure}
\includegraphics[angle=90.0,scale=0.65]{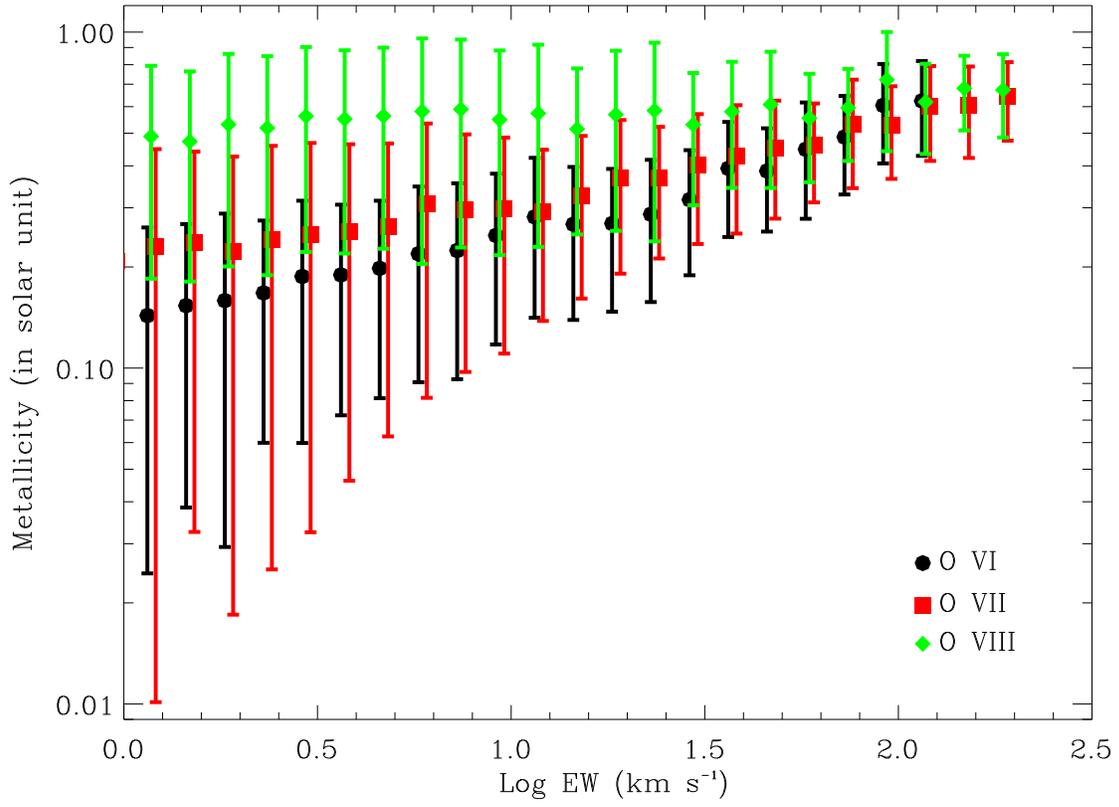}
\caption{
shows the average metallicity of O VI, O VI, O VIII absorption 
lines as a function of equivalent widths.
The errorbars indicate the scatter about the average.
}
\label{f8}
\end{figure}

\begin{figure}
\includegraphics[angle=0.0,scale=0.65]{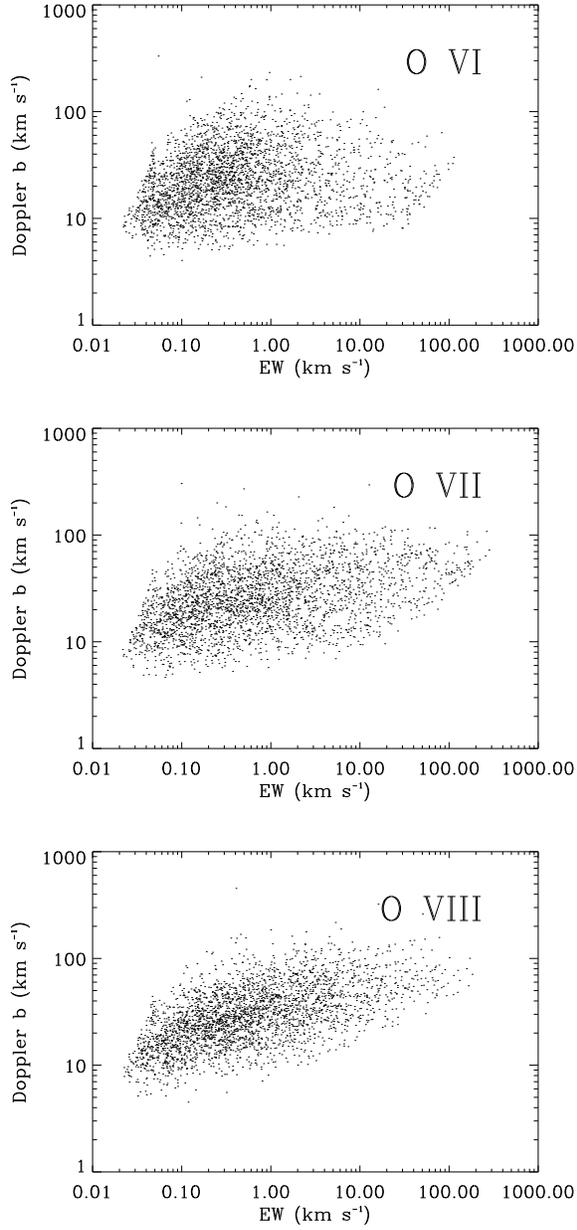}
\caption{
shows the b-parameter of lines versus their equivalent widths,
for O VI (top), O VII (middle) and O VIII (bottom panel).
}
\label{f8}
\end{figure}

\clearpage
\begin{figure}
\includegraphics[angle=90.0,scale=0.55]{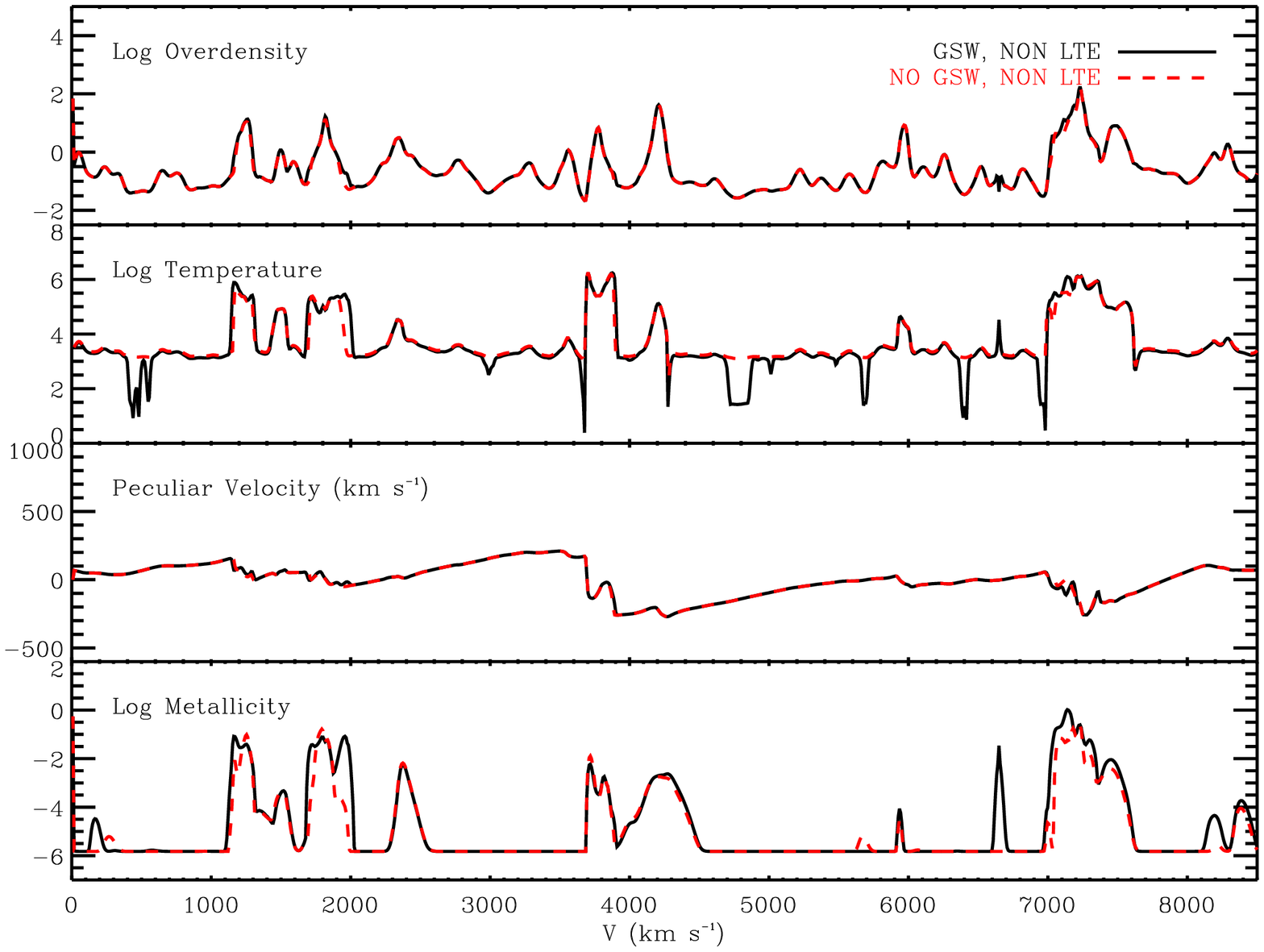}
\includegraphics[angle=90.0,scale=0.55]{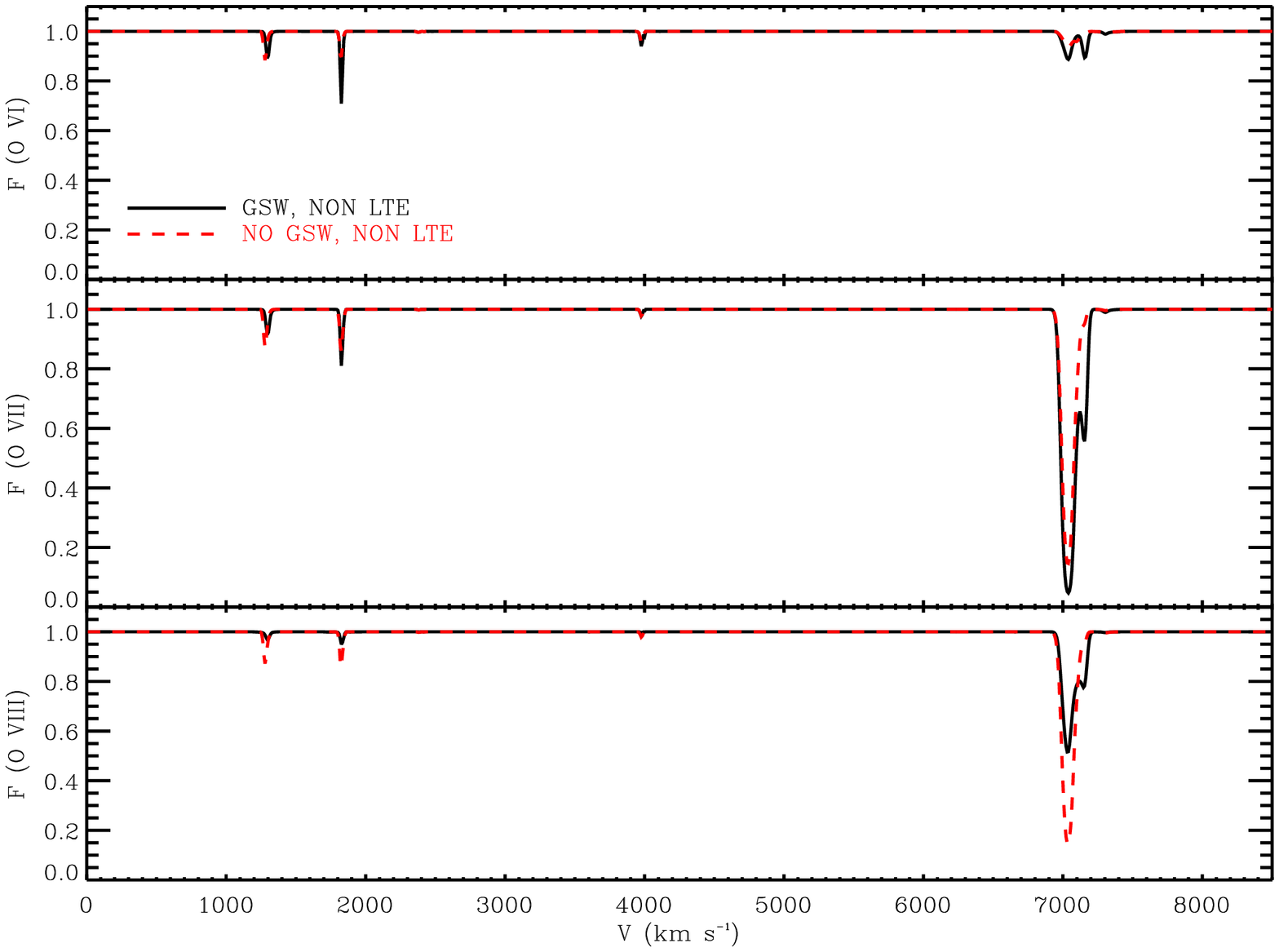}
\caption{
shows one line of sight through the simulation box.
The top four panels from top to bottom give
gas overdensity, temperature, peculiar velocity and 
metallicity (in solar units), respectively.
The bottom three panels from top to bottom are
flux absorption spectra for O VI, O VII and O VIII lines,
respectively. The solid curves are for the simulation
with GSW and the dashed curves without GSW.
}
\label{f9}
\end{figure}

\clearpage
\begin{figure}
\includegraphics[angle=90.0,scale=0.55]{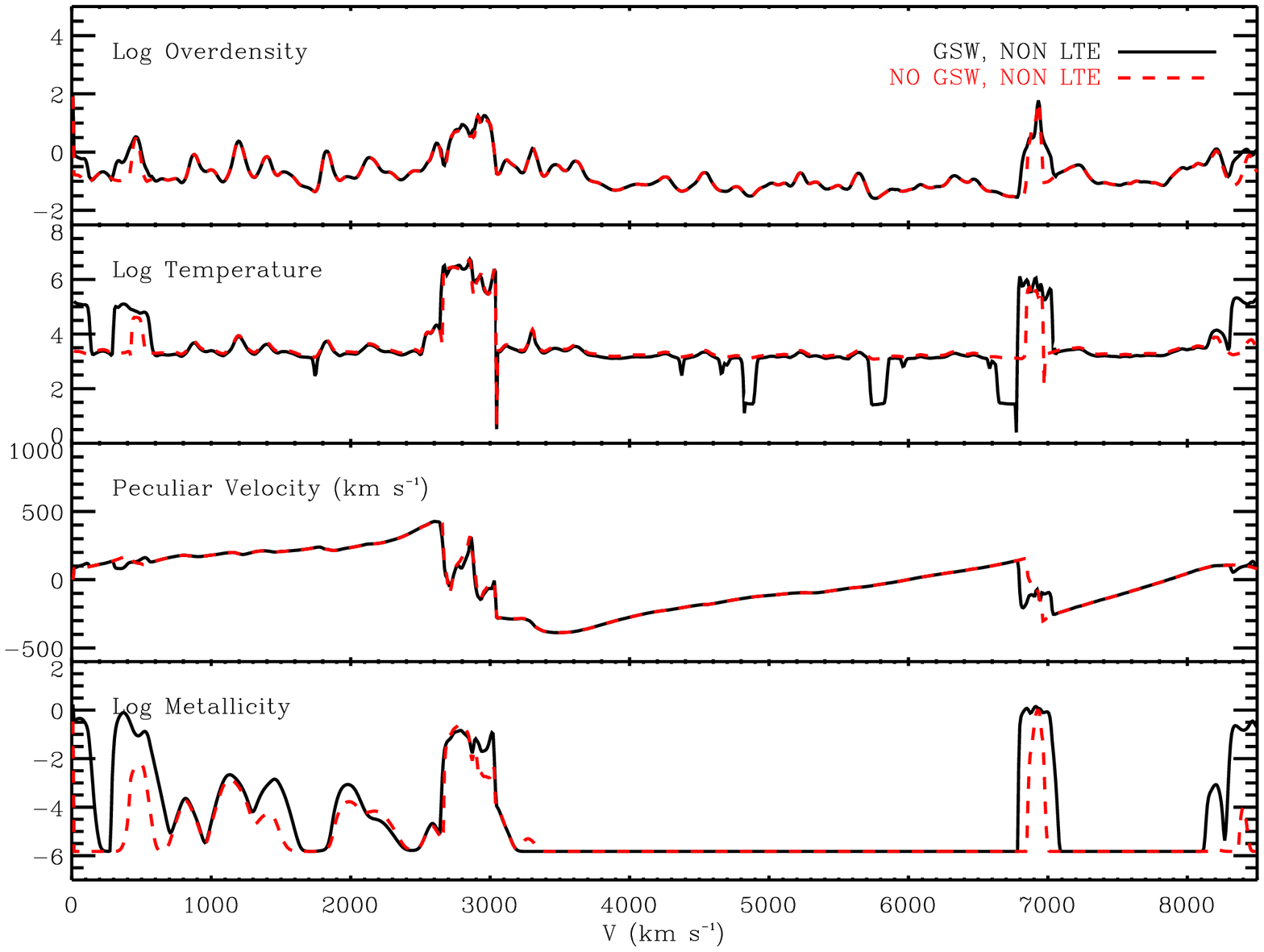}
\includegraphics[angle=90.0,scale=0.55]{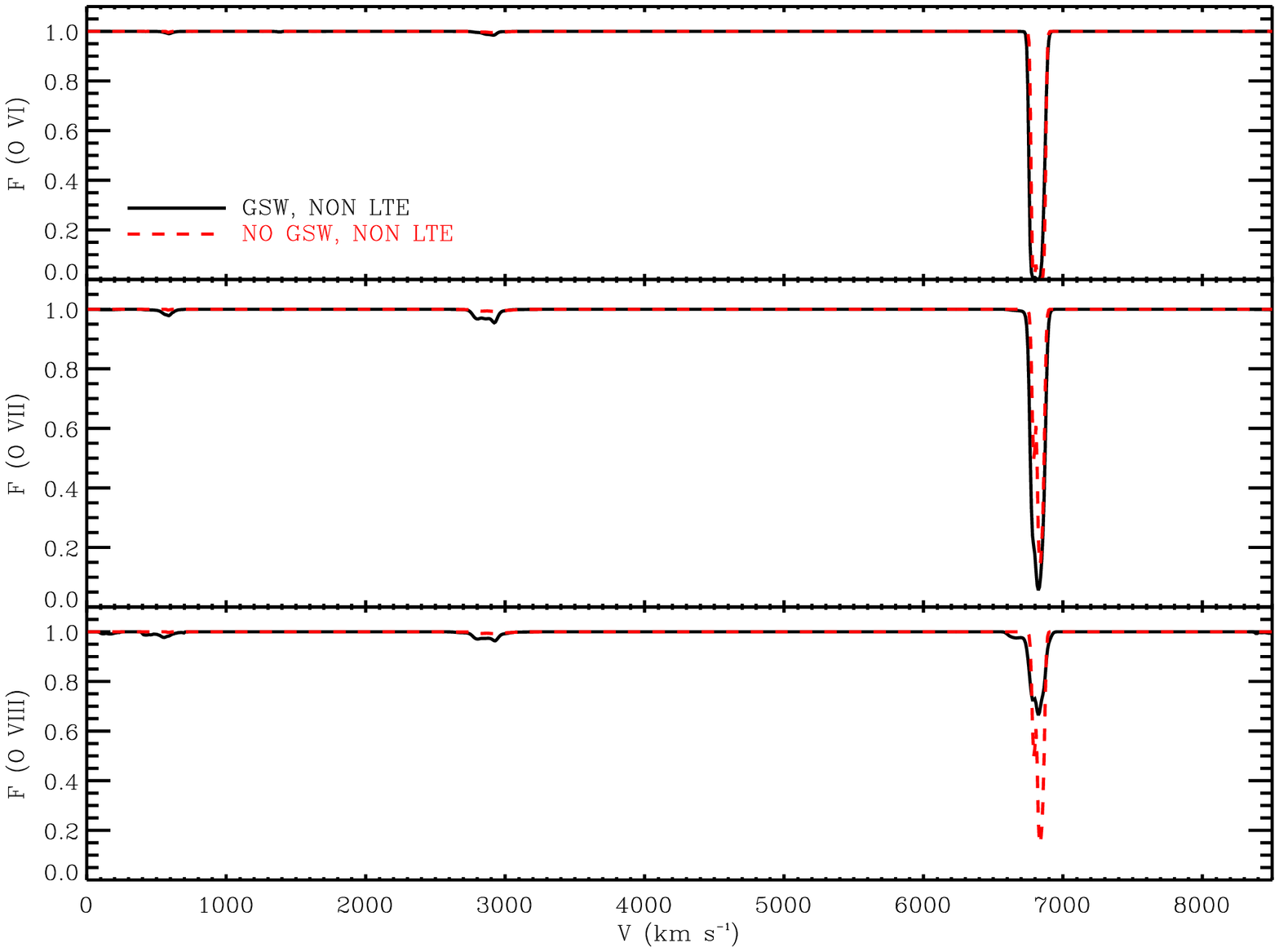}
\caption{
shows another line of sight through the simulation box.
The top four panels from top to bottom give
gas overdensity, temperature, peculiar velocity and 
metallicity (in solar units), respectively.
The bottom three panels from top to bottom are
flux absorption spectra for O VI, O VII and O VIII lines,
respectively. The solid curves are for the simulation
with GSW and the dashed curves without GSW.
}
\label{f9}
\end{figure}

\clearpage
\begin{figure}
\includegraphics[angle=90.0,scale=0.55]{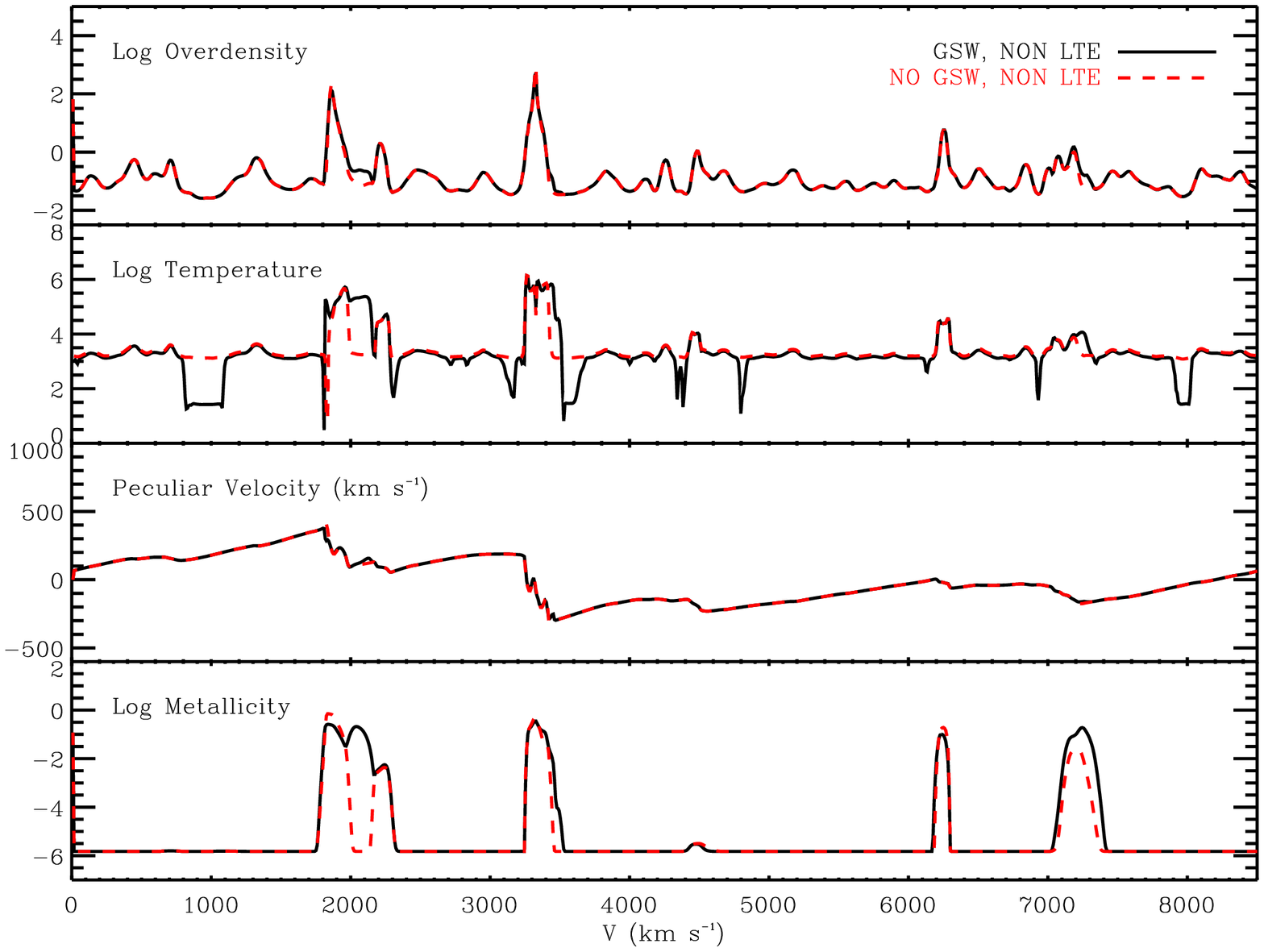}
\includegraphics[angle=90.0,scale=0.55]{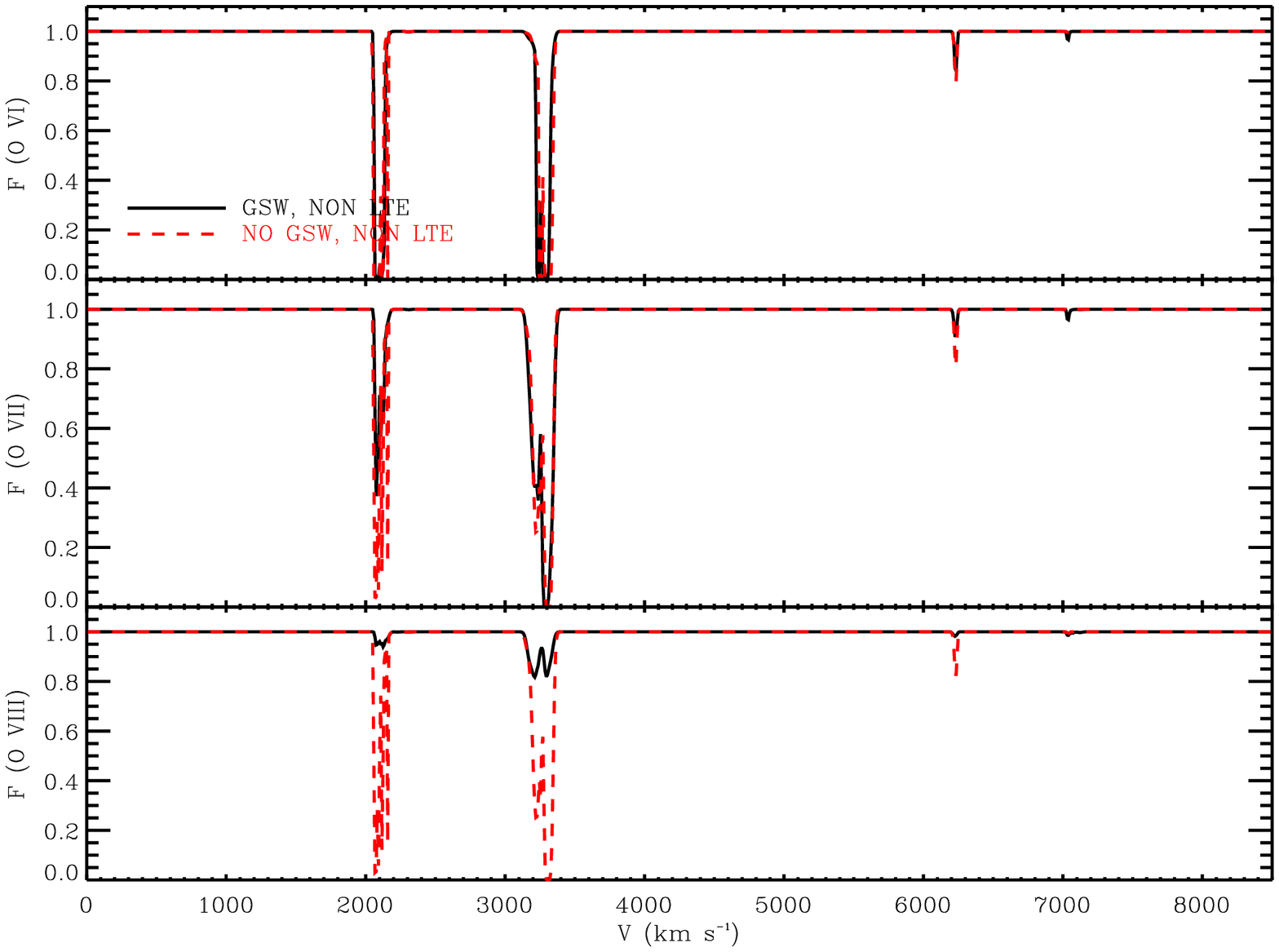}
\caption{
shows another line of sight through the simulation box.
The top four panels from top to bottom give
gas overdensity, temperature, peculiar velocity and 
metallicity (in solar units), respectively.
The bottom three panels from top to bottom are
flux absorption spectra for O VI, O VII and O VIII lines,
respectively. The solid curves are for the simulation
with GSW and the dashed curves without GSW.
}
\label{f9}
\end{figure}

\begin{figure}
\includegraphics[angle=0.0,scale=0.80]{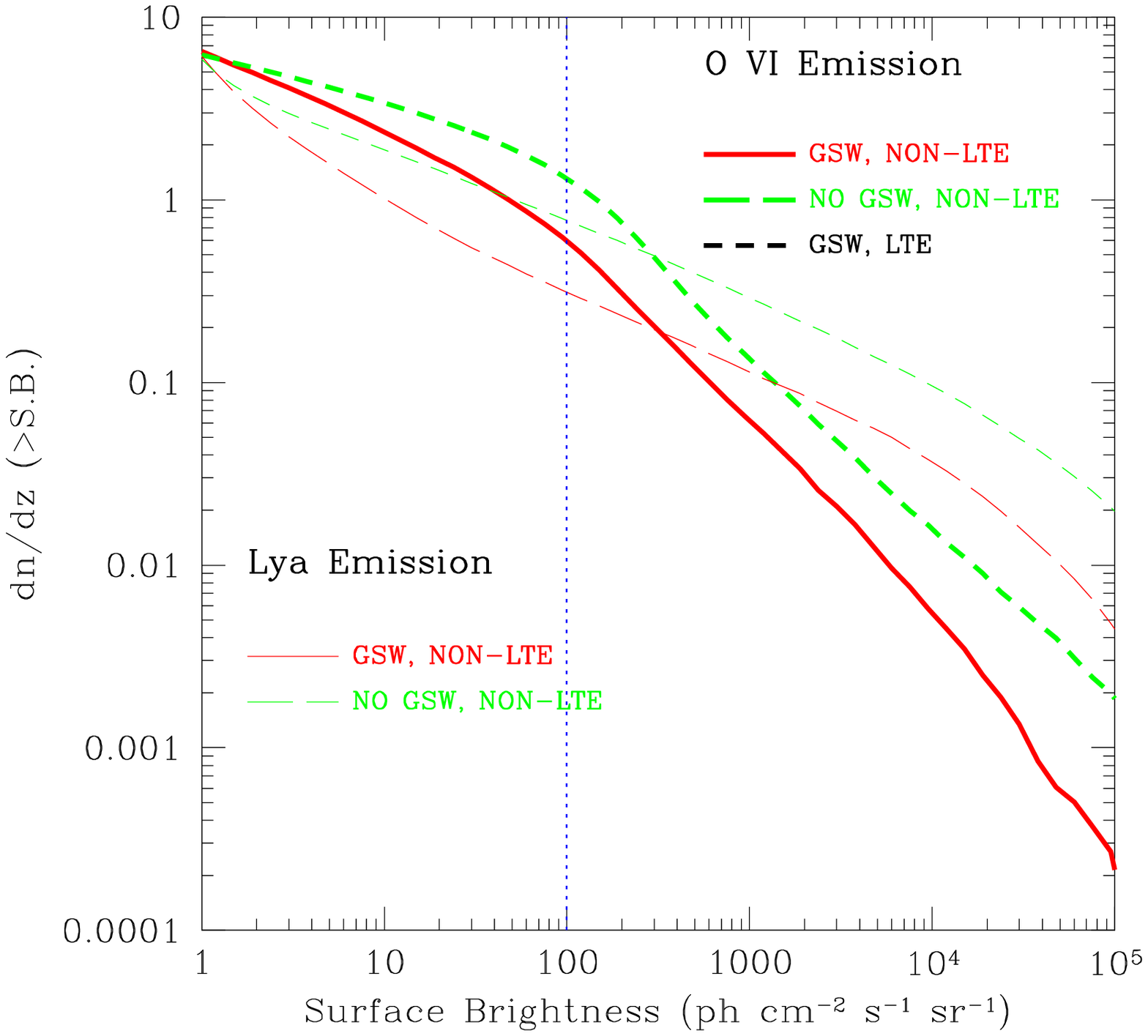}
\caption{
shows the cumulative number of O VI (thick curves) and \lya lines (thin curves)
per unit redshift as a function of surface brightness.
The red curves show results from the simulation
with GSW and the green curve from the simulation without GSW.
The black curve (for O VI line only) is computed using CLOUDY code
on the assumption of ionization equilibrium
based on the density, temperature
and metallicity information from the simulation with GSW.
Note that future planned UV missions may be able
to achieve a sensitivity in the range $100$ in the displayed units,
as indicated by the vertical blue line.
}
\label{f10}
\end{figure}

\begin{figure}
\includegraphics[angle=0.0,scale=0.80]{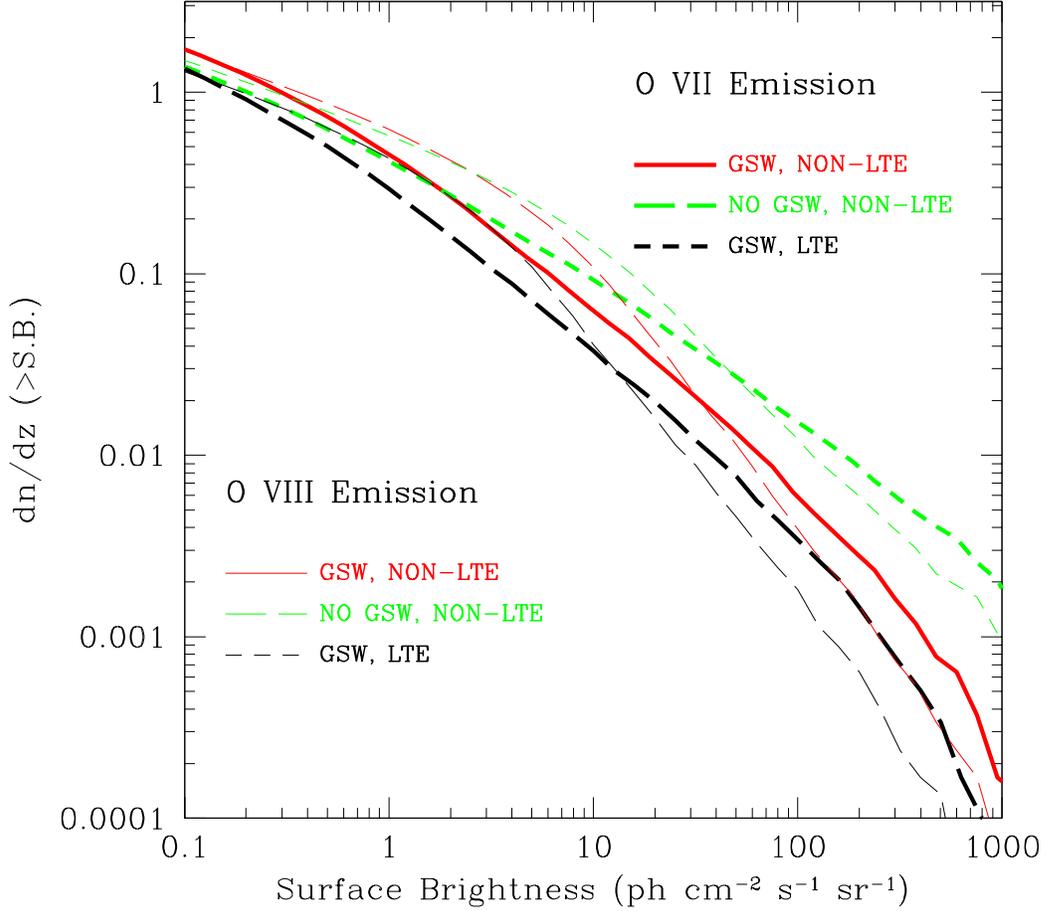}
\caption{
shows the cumulative number of O VII (thick curves) and O VIII lines 
(thin curves)
per unit redshift as a function of surface brightness.
The red curves show results from the simulation
with GSW and the blue curve from the simulation without GSW.
The green curve is computed using CLOUDY code
on the assumption of ionization equilibrium
based on the density, temperature
and metallicity information from the simulation with GSW.
The planned Japanese soft X-ray emission DIOS (Yoshikawa \etal 2003)
will have a sensitivity of order $0.1$ in the displayed units.
}
\label{f5}
\end{figure}

\begin{figure}
\includegraphics[angle=0.0,scale=0.80]{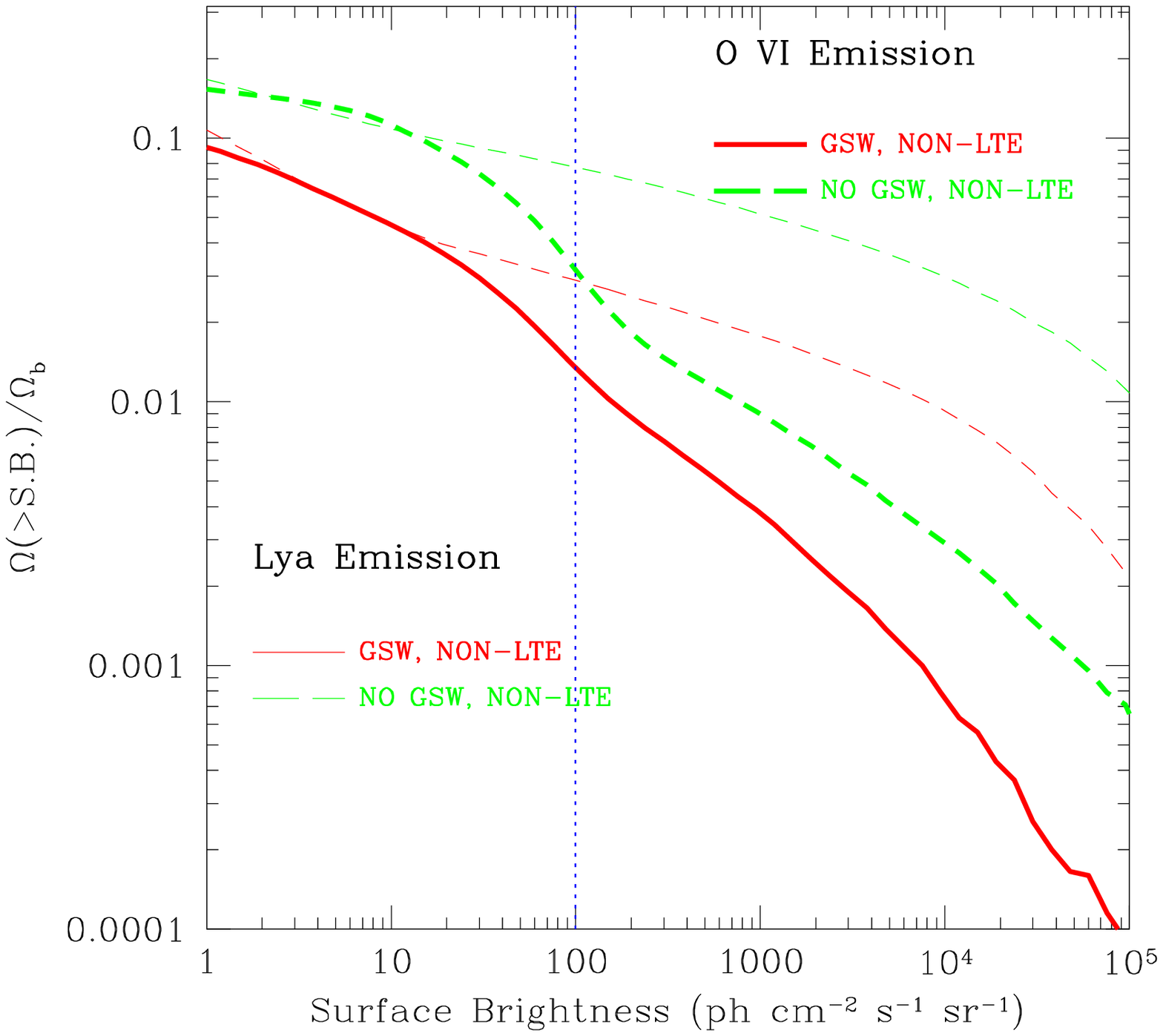}
\caption{
shows the cumulative gas mass density probed by 
the O VI (thick curves) and \lya (thin curves) lines as a function of surface brightness.
The red curves show results from the simulation
with GSW and the green curves from the simulation without GSW.
}
\label{f10}
\end{figure}

\begin{figure}
\includegraphics[angle=0.0,scale=0.80]{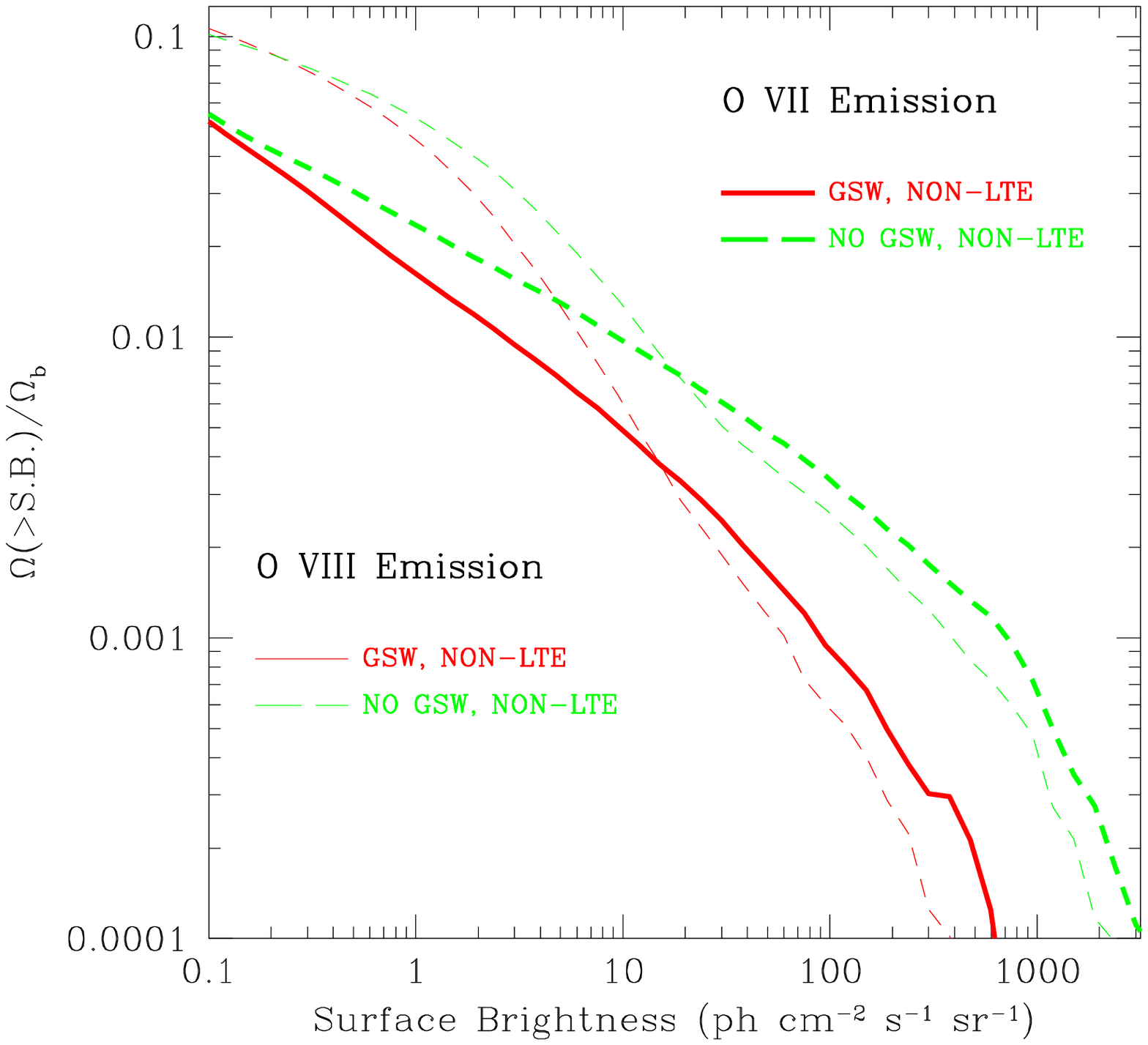}
\caption{
shows the cumulative gas mass density probed by 
the O VII (thick curves) and O VIII (thin curves) 
lines as a function of surface brightness.
The red curves show results from the simulation
with GSW and the green curves from the simulation without GSW.
}
\label{f5}
\end{figure}

\begin{figure}
\includegraphics[angle=0.0,scale=0.80]{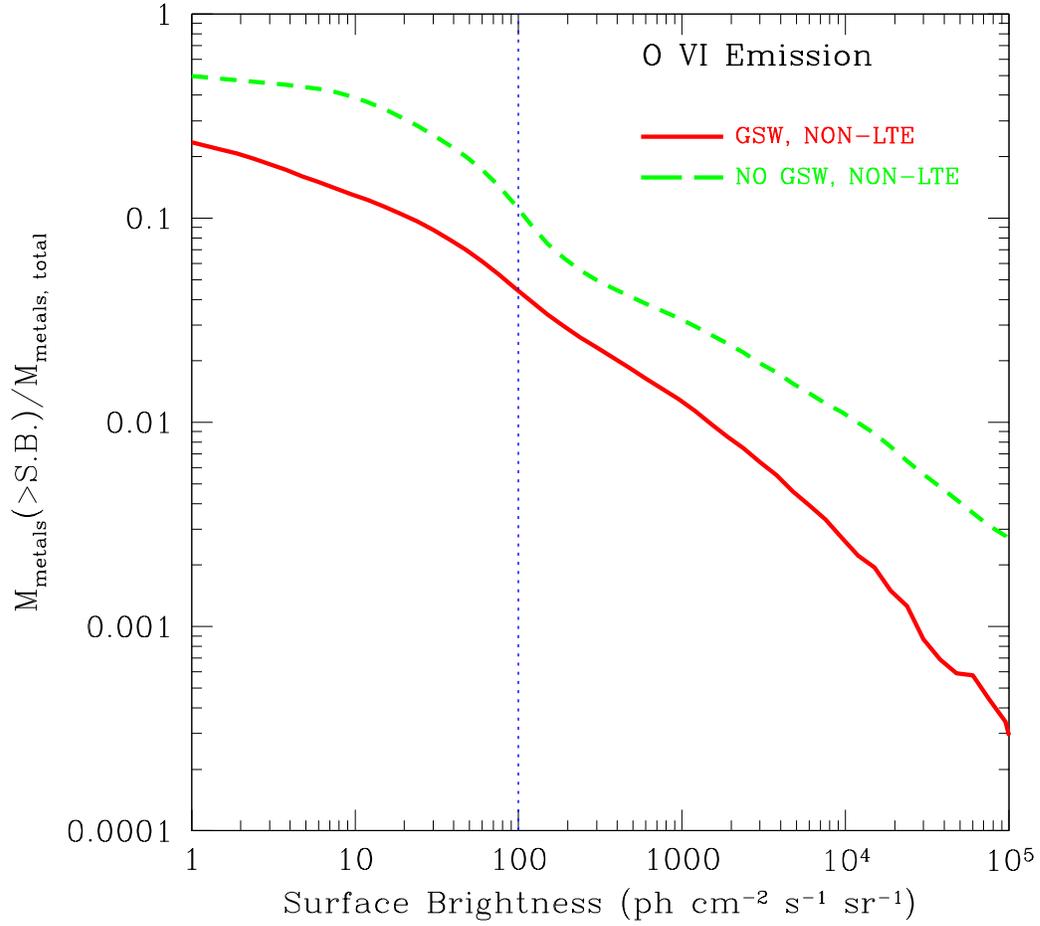}
\caption{
shows the cumulative metal mass probed by 
the O VI line as a function of surface brightness.
The red curves show results from the simulation
with GSW and the green curves from the simulation without GSW.
}
\label{f10}
\end{figure}

\begin{figure}
\includegraphics[angle=0.0,scale=0.80]{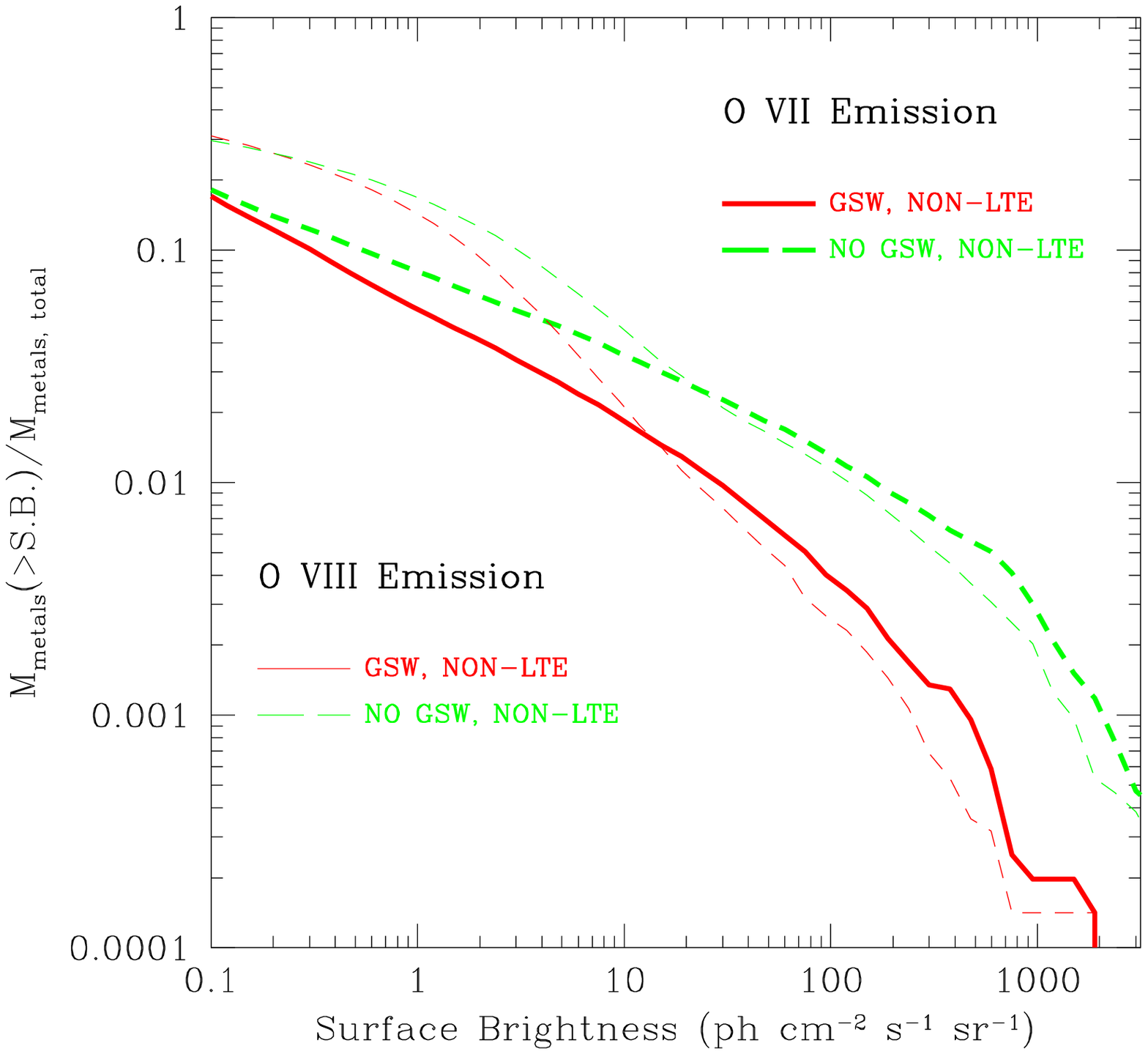}
\caption{
shows the cumulative metal mass probed by 
the O VII (thick curves) and O VIII (thin curves) 
lines as a function of surface brightness.
The red curves show results from the simulation
with GSW and the green curves from the simulation without GSW.
}
\label{f5}
\end{figure}

\clearpage 
\begin{figure}
\includegraphics[angle=0.0,scale=0.80]{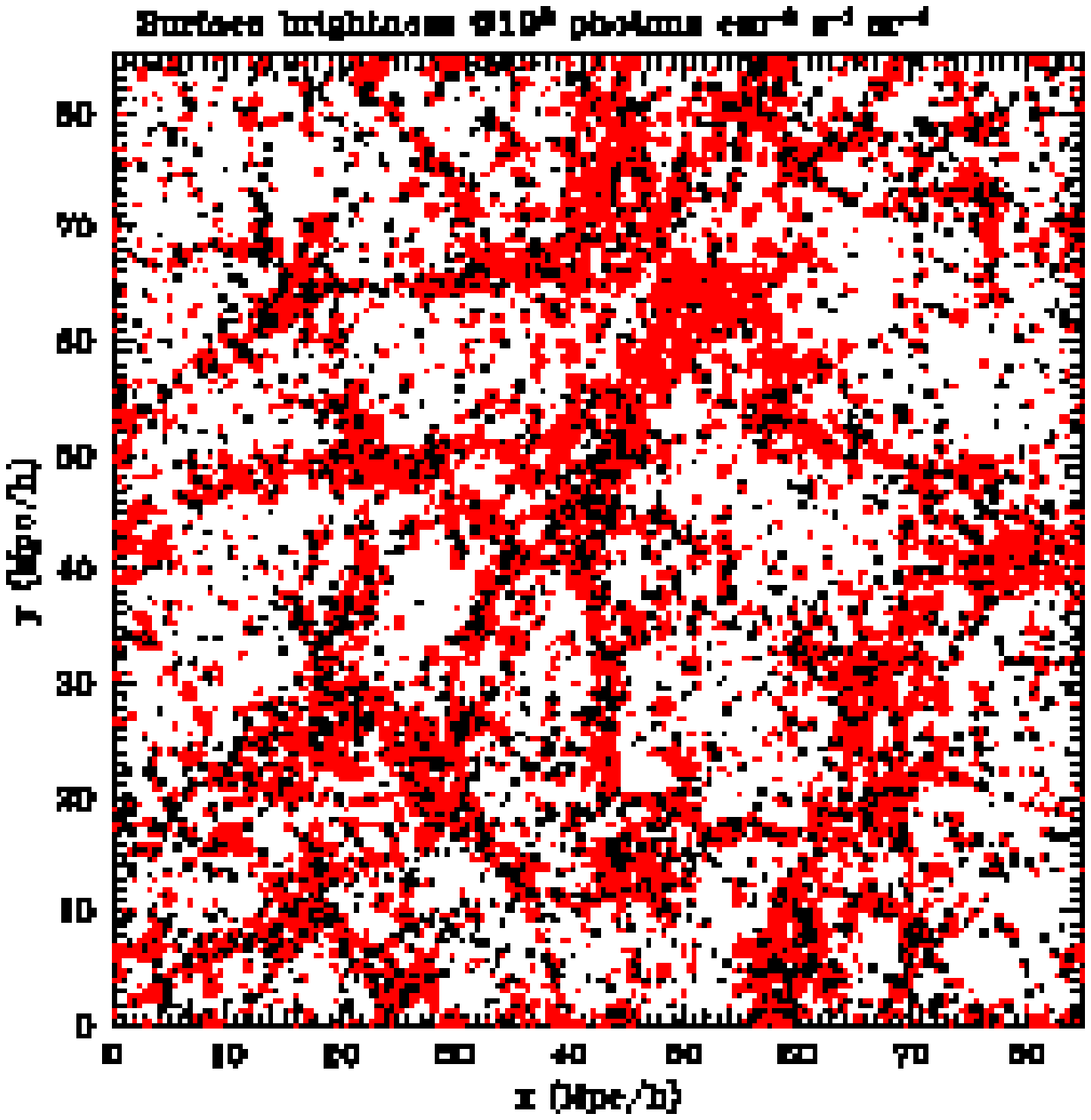}
\caption{
shows a map of size
$85\times 85$Mpc$^2$/h$^2$ with a depth of $85$Mpc/h,
with the observable regions shown as black spots
for O VI emission line
with an instrument of a sensitivity of 100 photon~cm$^{-2}$sr$^{-1}$s$^{-1}$.
The red contours are the underlying gas density distribution.
}
\label{f5}
\end{figure}

\clearpage 
\begin{figure}
\includegraphics[angle=0.0,scale=0.80]{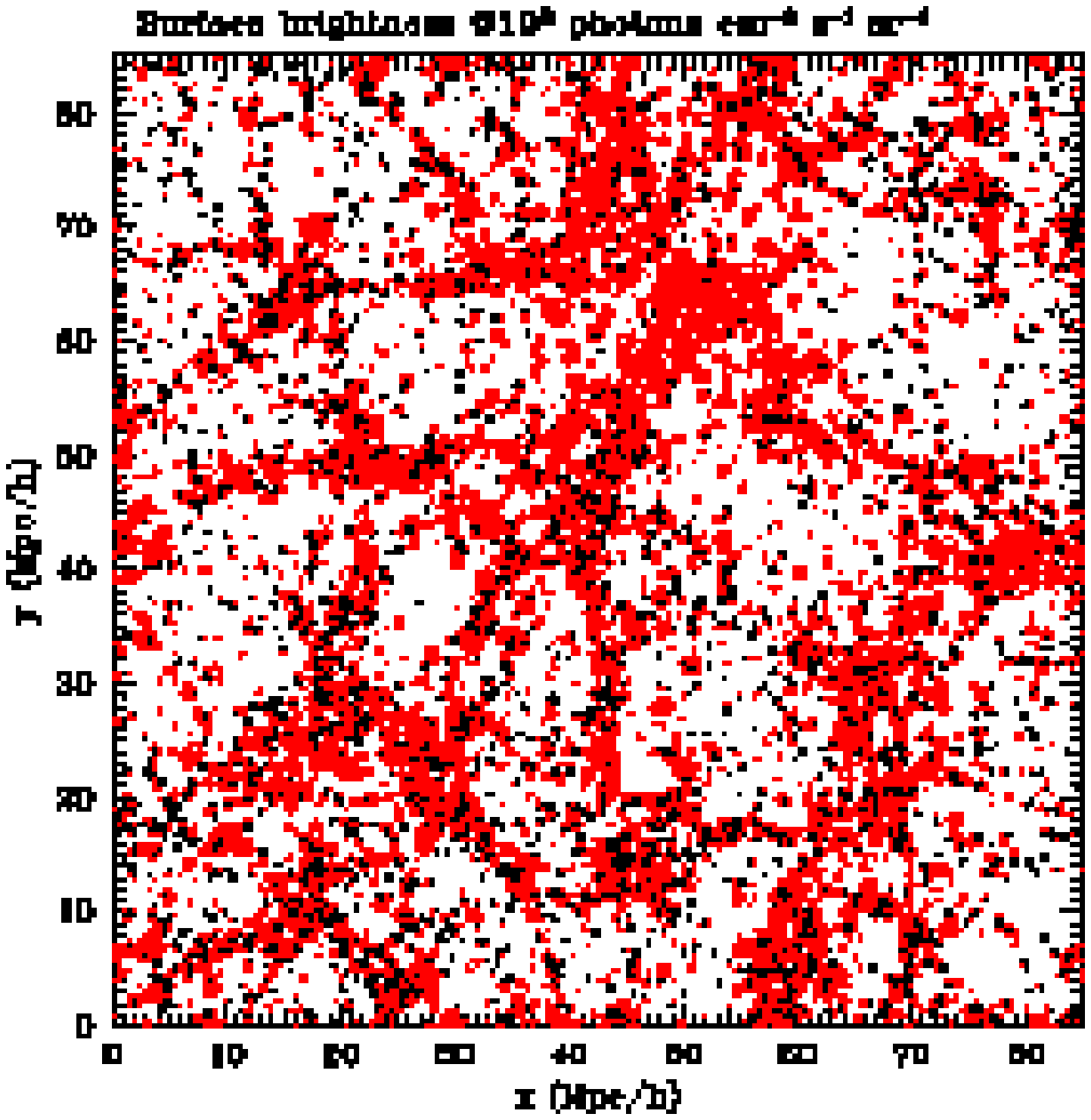}
\caption{
shows a map of size
$85\times 85$Mpc$^2$/h$^2$ with a depth of $85$Mpc/h,
with the observable regions shown as black spots
for \lya emission line
with an instrument of a sensitivity of 100 photon~cm$^{-2}$sr$^{-1}$s$^{-1}$.
The red contours are the underlying gas density distribution.
}
\label{f5}
\end{figure}

\clearpage 
\begin{figure}
\includegraphics[angle=0.0,scale=0.80]{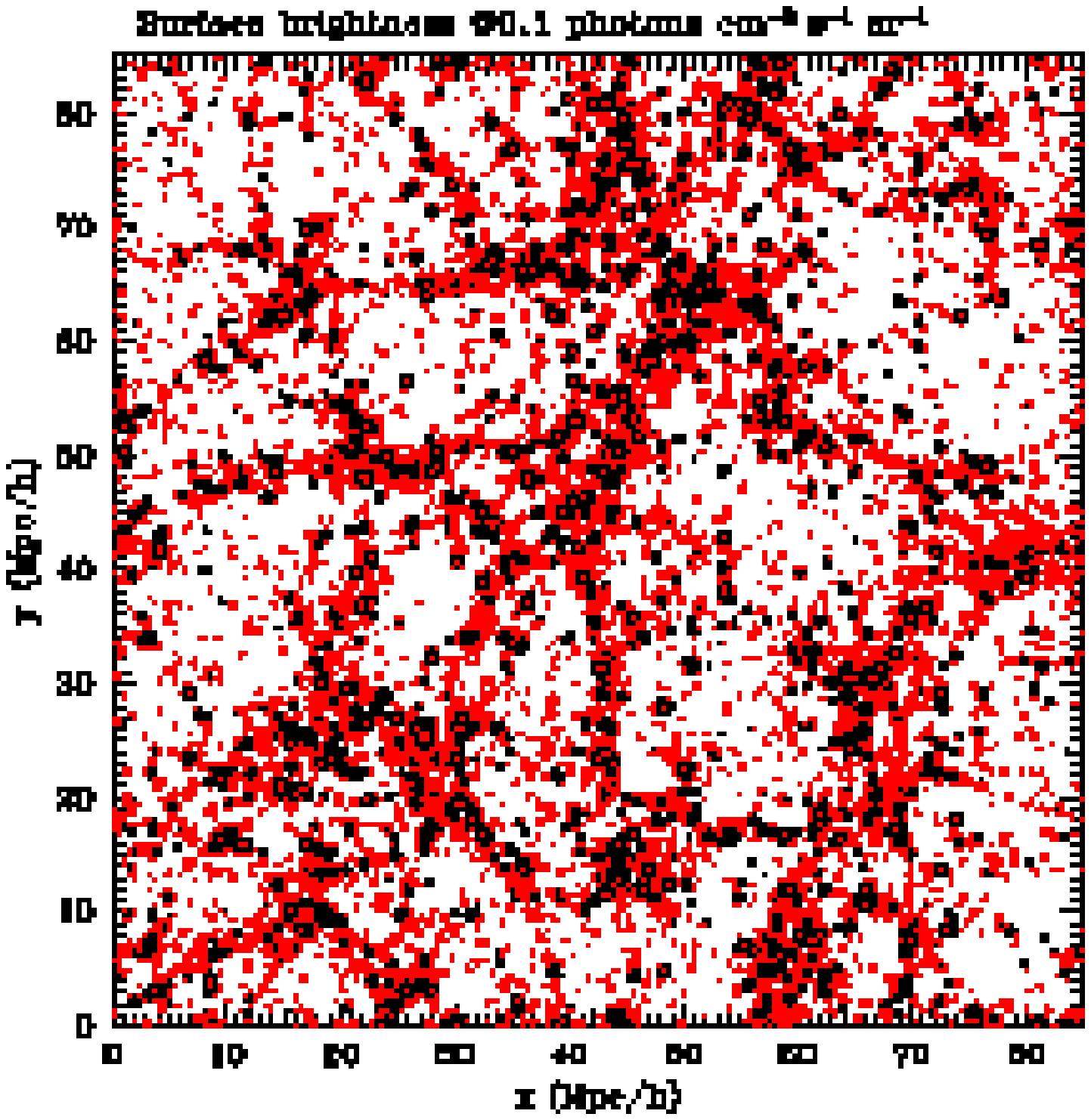}
\caption{
shows a map of size
$85\times 85$Mpc$^2$/h$^2$ with a depth of $85$Mpc/h,
with the observable regions shown as black spots
for O VII emission line
with an instrument of a sensitivity of 0.1~photon~cm$^{-2}$sr$^{-1}$s$^{-1}$.
The red contours are the underlying gas density distribution.
}
\label{f5}
\end{figure}

\clearpage 
\begin{figure}
\includegraphics[angle=0.0,scale=0.80]{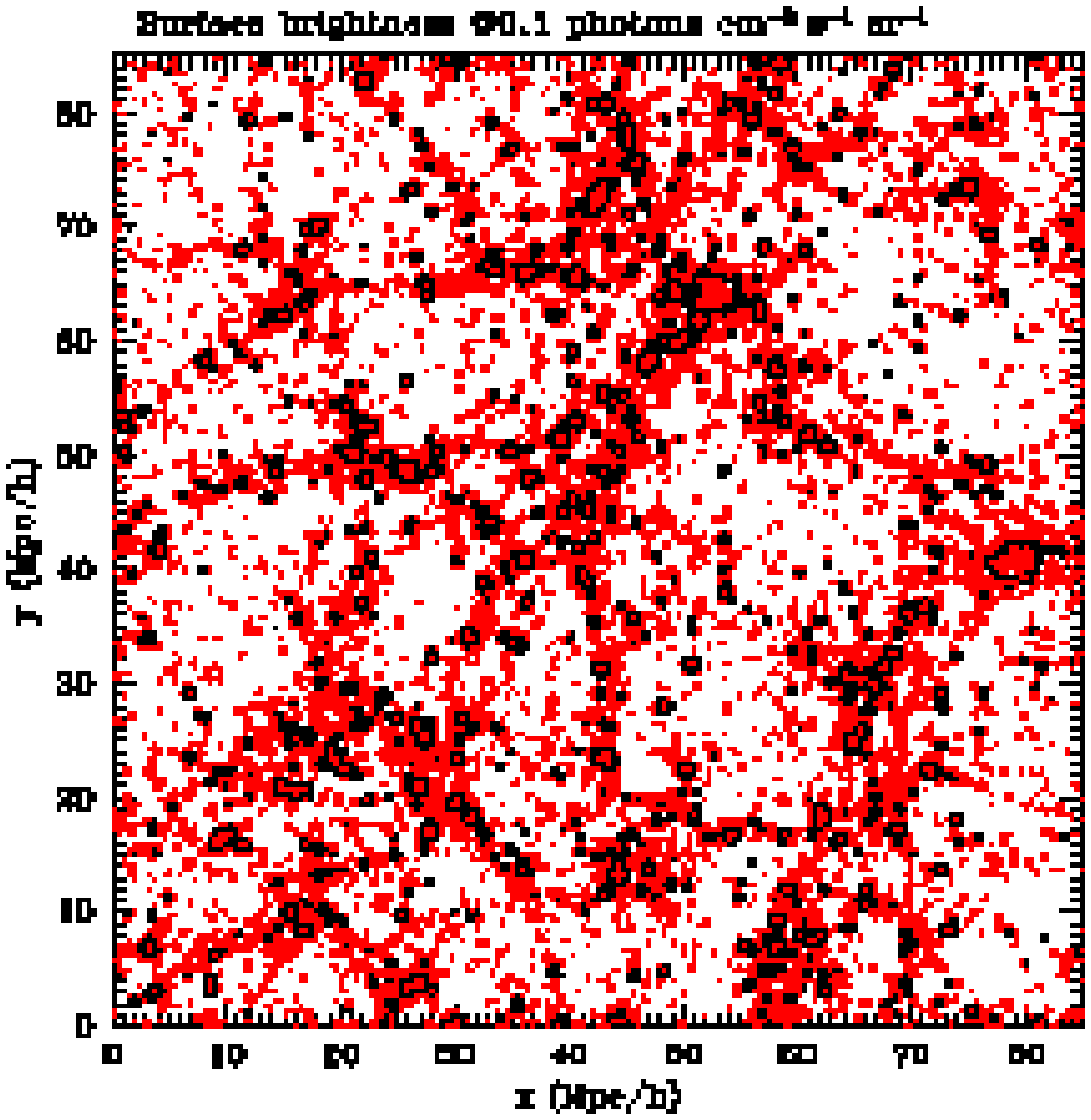}
\caption{
shows a map of size
$85\times 85$Mpc$^2$/h$^2$ with a depth of $85$Mpc/h,
with the observable regions shown as black spots
for O VIII emission line
with an instrument of a sensitivity of 0.1 photon~cm$^{-2}$sr$^{-1}$s$^{-1}$.
The red contours are the underlying gas density distribution.
}
\label{f5}
\end{figure}

\clearpage 
\begin{figure}
\includegraphics[angle=0.0,scale=0.80]{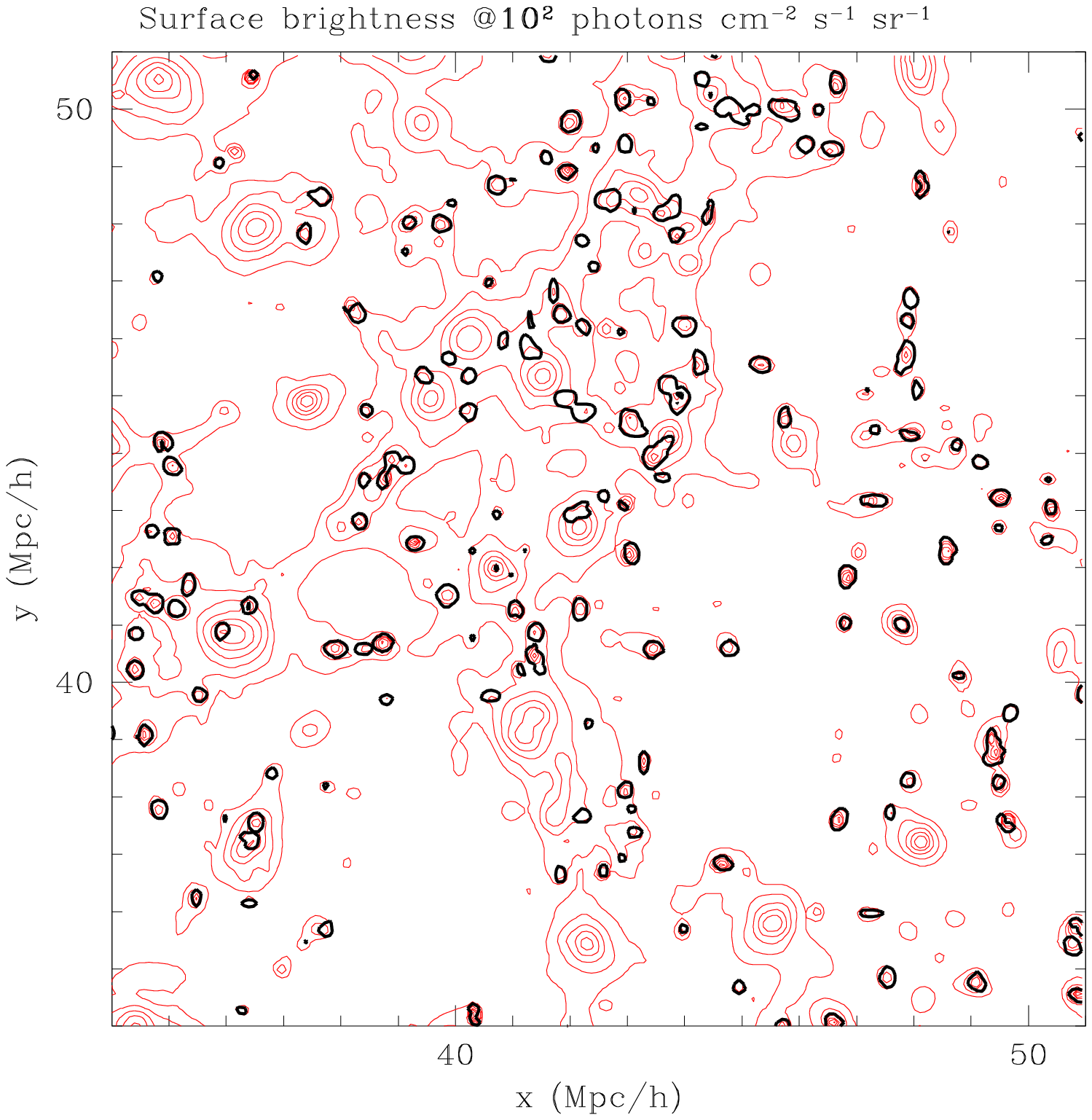}
\caption{
shows a part of Figure 17 by
zooming into a small region to better display some details,
showing the observable regions shown as black spots
with O VI emission line
with an instrument of a sensitivity of 100 photon~cm$^{-2}$sr$^{-1}$s$^{-1}$.
The red contours are the underlying gas density distribution.
}
\label{f16}
\end{figure}

\clearpage 
\begin{figure}
\includegraphics[angle=0.0,scale=0.80]{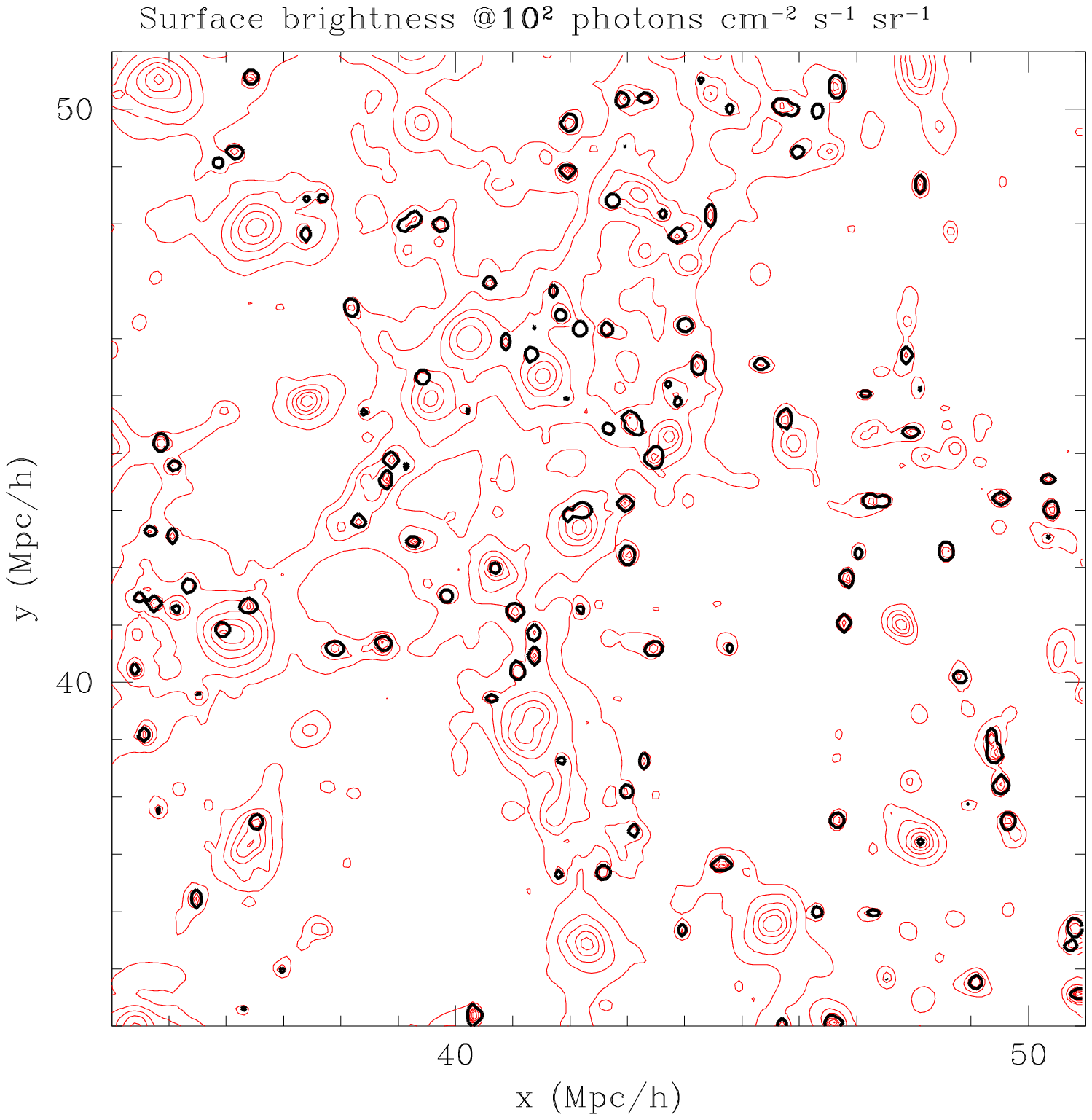}
\caption{
shows a part of Figure 18 by
zooming into a small region to better display some details,
showing the observable regions shown as black spots
with \lya emission line
with an instrument of a sensitivity of 100 photon~cm$^{-2}$sr$^{-1}$s$^{-1}$.
The red contours are the underlying gas density distribution.
}
\label{f16}
\end{figure}

\clearpage 
\begin{figure}
\includegraphics[angle=0.0,scale=0.80]{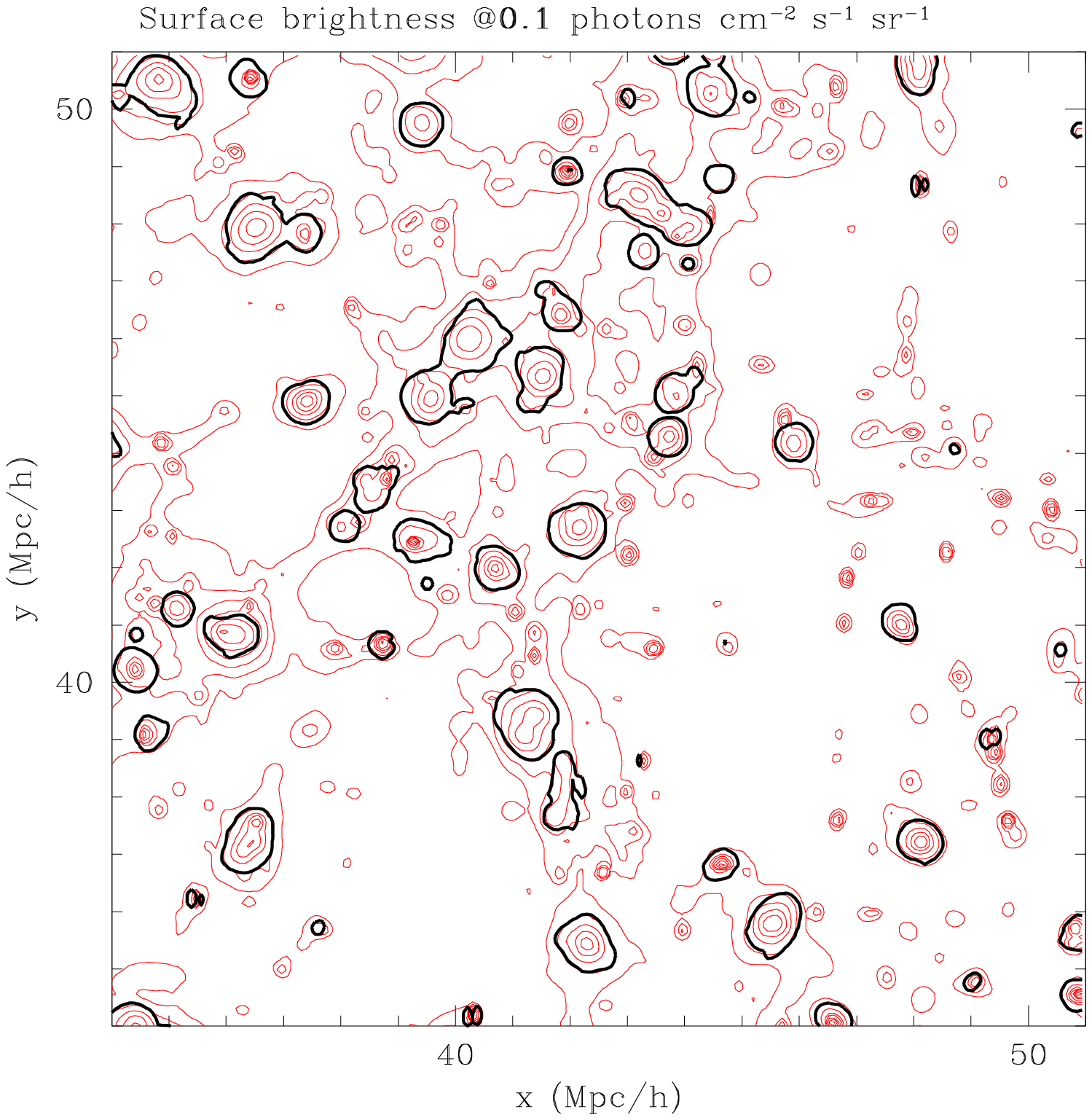}
\caption{
shows a part of Figure 19 by
zooming into a small region to better display some details,
showing the observable regions shown as black spots
with O VII emission line
with an instrument of a sensitivity of 0.1~photon~cm$^{-2}$sr$^{-1}$s$^{-1}$.
The red contours are the underlying gas density distribution.
}
\label{f16}
\end{figure}

\clearpage 
\begin{figure}
\includegraphics[angle=0.0,scale=0.80]{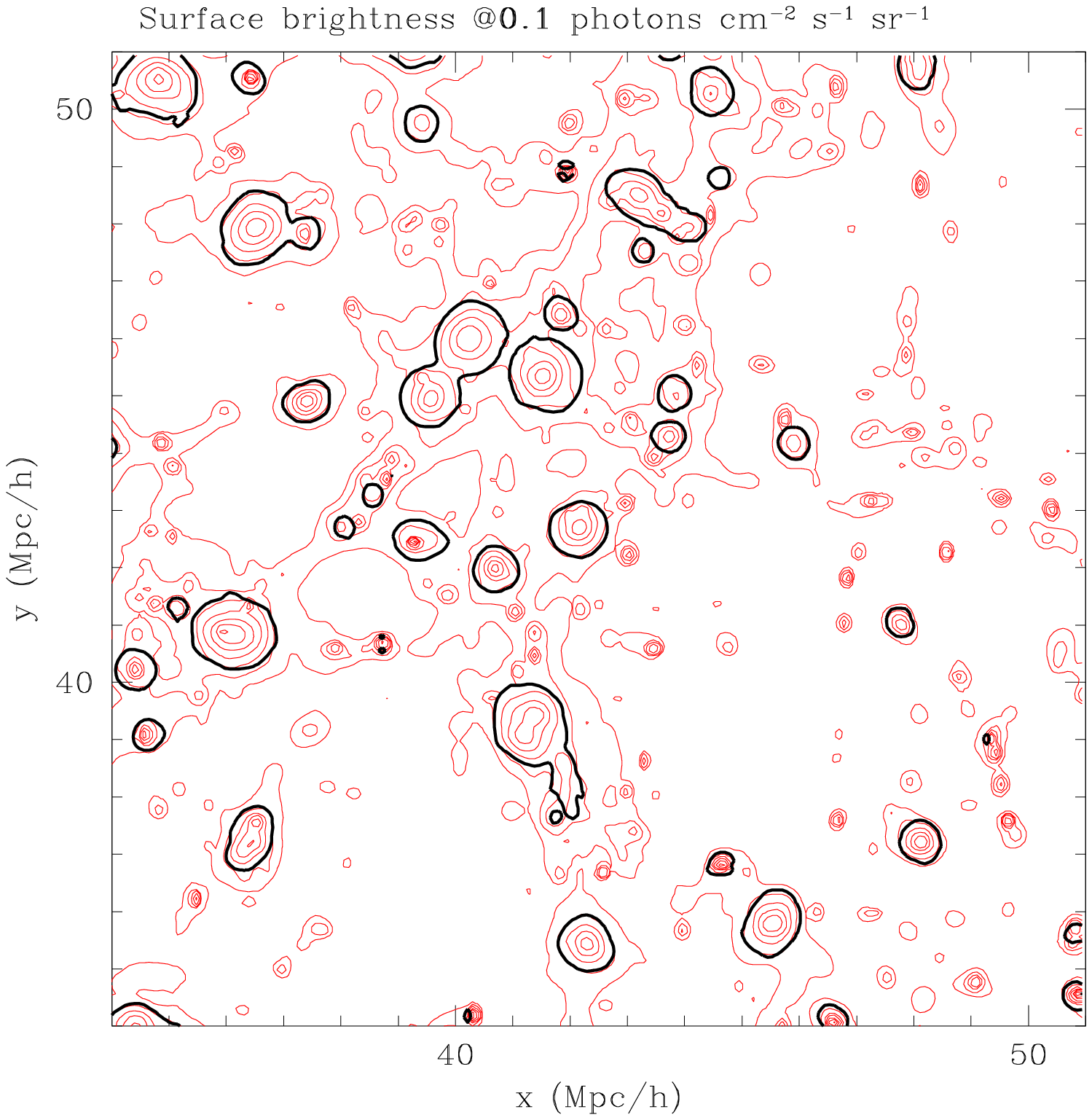}
\caption{
shows a part of Figure 20 by
zooming into a small region to better display some details,
showing the observable regions shown as black spots
with O VIII emission line
with an instrument of a sensitivity of 0.1~photon~cm$^{-2}$sr$^{-1}$s$^{-1}$.
The red contours are the underlying gas density distribution.
}
\label{f16}
\end{figure}

\begin{figure}
\includegraphics[angle=90.0,scale=0.70]{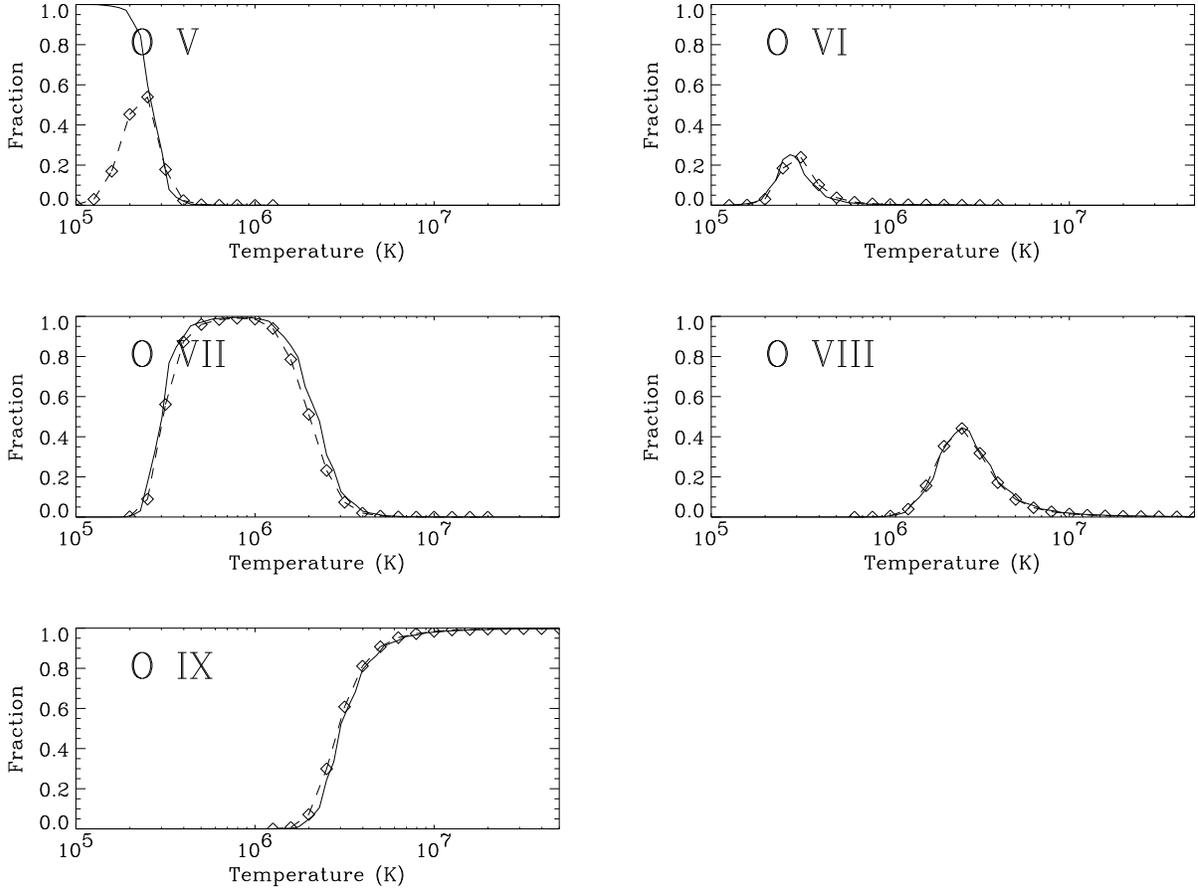}
\caption{
shows the comparison between our calculation with very long integration time 
and equilibrium calculation for the collisional ionization case
with zero background radiation field. 
The solid lines in each
panel are results from our calculation, and the diamonds are adopted
from Mazzotta et al.~(1998), based on collisional ionization equilibrium. 
}
\label{f1}
\end{figure}

\begin{figure}
\includegraphics[angle=90.0,scale=0.70]{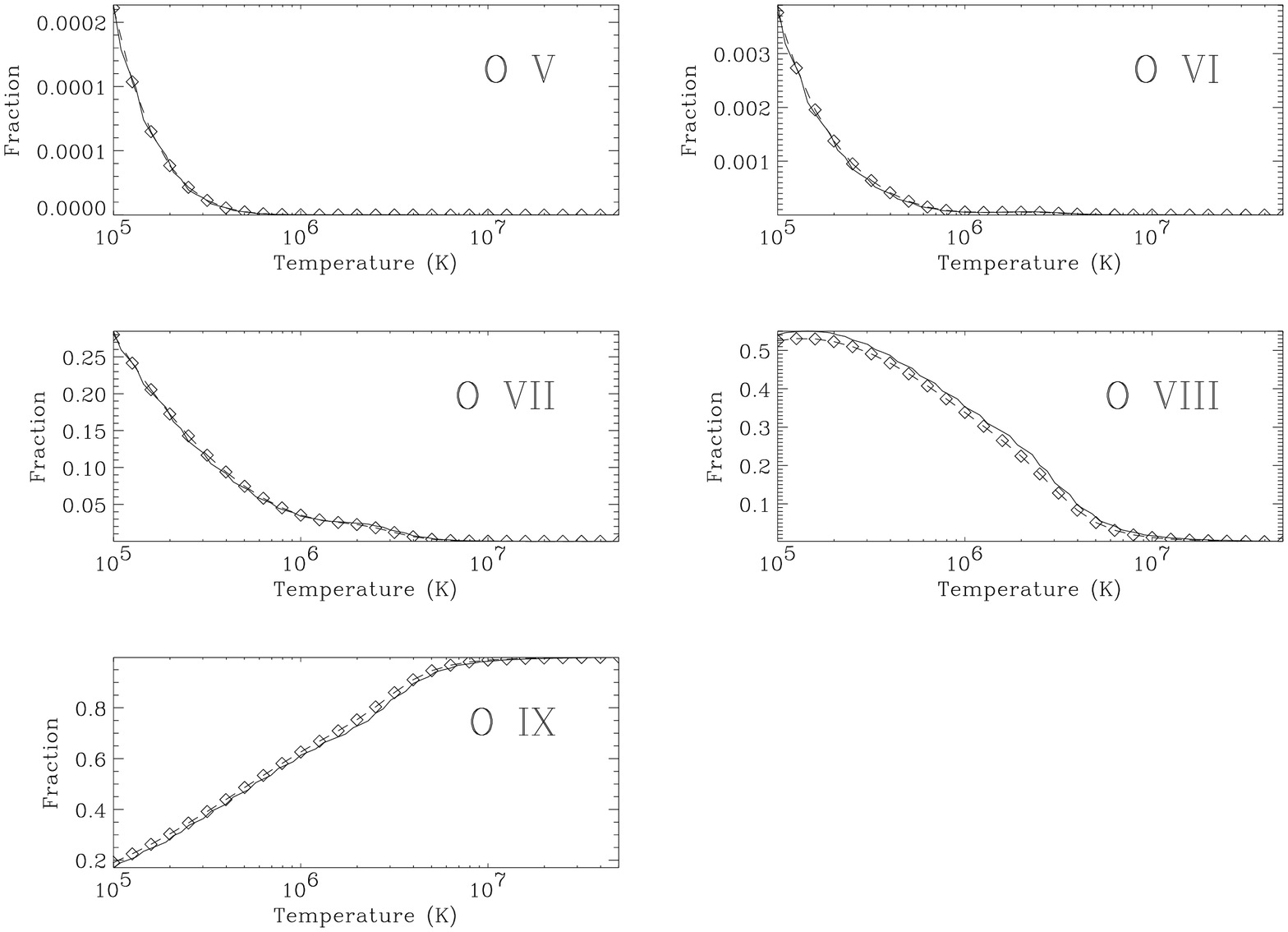}
\caption{
a similar comparison between our calculation as in Figure 25,
with an addition of a background radiation 
field of $J(912\AA) = 10^{-22} \rm\ ergs\ s^{-1}cm^{-2}Hz^{-1}sr^{-1}$. 
}
\label{f1}
\end{figure}

\begin{figure}
\includegraphics[angle=90.0,scale=0.70]{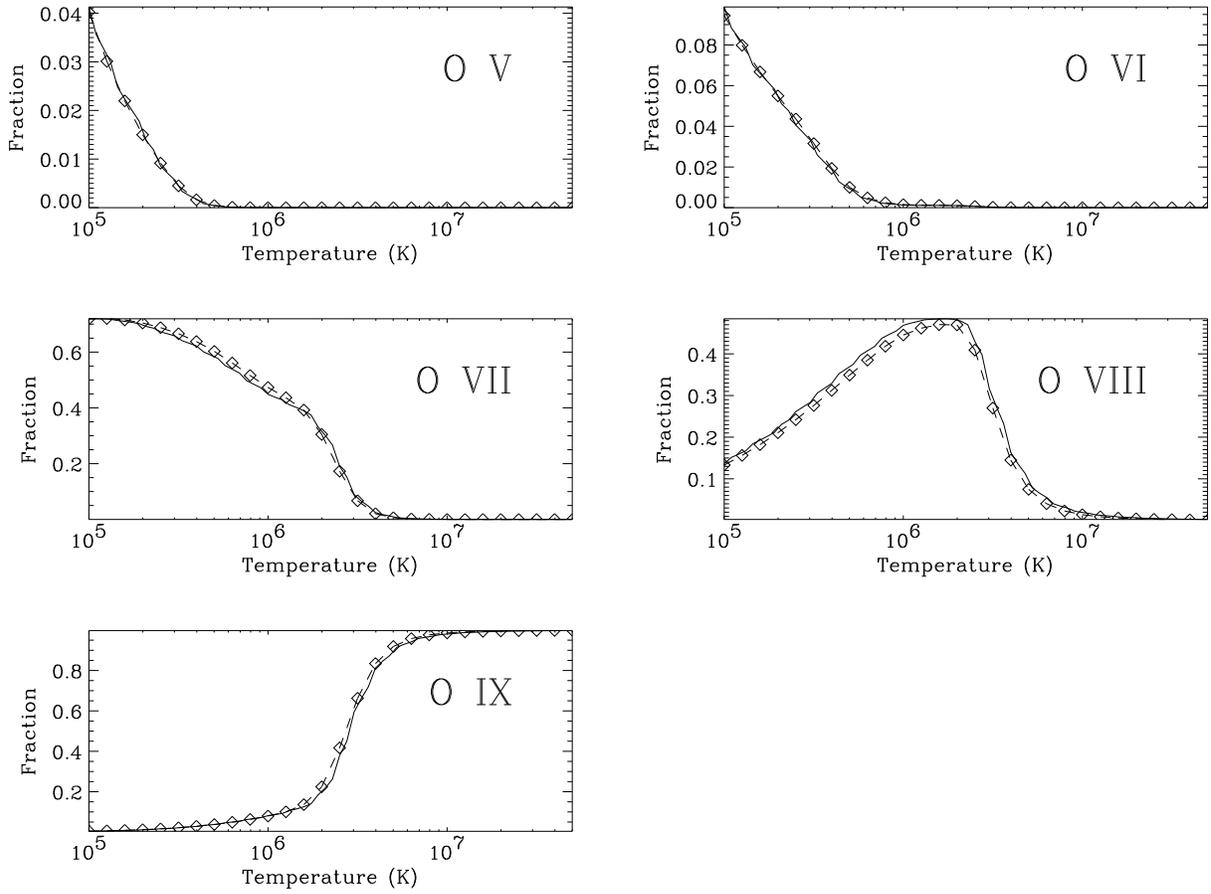}
\caption{
a similar comparison between our calculation as in Figure 25,
with a different background radiation
field of $J(912\AA) = 10^{-23} \rm\ ergs\ s^{-1}cm^{-2}Hz^{-1}sr^{-1}$. 
}
\label{f1}
\end{figure}

\end{document}